\documentclass[12pt]{article}

\usepackage[utf8]{inputenc}
\usepackage[T1]{fontenc}
\usepackage{lmodern}
\usepackage{amsmath,amssymb,amsthm}
\usepackage{mathtools}
\usepackage{textcomp}
\usepackage{graphicx}
\usepackage{booktabs}
\usepackage{longtable}
\usepackage{etoolbox}
\AtBeginEnvironment{longtable}{\small}
\usepackage{tabularx}
\usepackage{array}
\usepackage{hyperref}
\usepackage{url}
\usepackage[authoryear,round]{natbib}
\usepackage[normalem]{ulem}
\usepackage[margin=1in]{geometry}
\usepackage{listings}
\usepackage{xcolor}
\usepackage{adjustbox}
\usepackage{fvextra}
\fvset{breaklines=true,breakanywhere=true,fontsize=\small}
  {\VerbatimEnvironment\begin{Verbatim}}%
  {\end{Verbatim}}

\usepackage{newunicodechar}
\newunicodechar{∀}{\ensuremath{\forall}}
\newunicodechar{∃}{\ensuremath{\exists}}
\newunicodechar{∈}{\ensuremath{\in}}
\newunicodechar{∪}{\ensuremath{\cup}}
\newunicodechar{∧}{\ensuremath{\wedge}}
\newunicodechar{→}{\ensuremath{\rightarrow}}
\newunicodechar{⇔}{\ensuremath{\Leftrightarrow}}
\newunicodechar{≠}{\ensuremath{\neq}}
\newunicodechar{≡}{\ensuremath{\equiv}}
\newunicodechar{≤}{\ensuremath{\leq}}
\newunicodechar{≥}{\ensuremath{\geq}}
\newunicodechar{−}{\ensuremath{-}}
\newunicodechar{±}{\ensuremath{\pm}}
\newunicodechar{§}{\S}
\newunicodechar{×}{\ensuremath{\times}}
\newunicodechar{⁺}{\ensuremath{^{+}}}
\newunicodechar{Δ}{\ensuremath{\Delta}}
\newunicodechar{ε}{\ensuremath{\varepsilon}}
\newunicodechar{○}{\ensuremath{\circ}}
\newunicodechar{●}{\ensuremath{\bullet}}
\newunicodechar{═}{=}
\newunicodechar{̂}{}

\providecommand{\llbracket}{\mathopen{\lbrack\!\lbrack}}
\providecommand{\rrbracket}{\mathclose{\rbrack\!\rbrack}}

\providecommand{\tightlist}{\setlength{\itemsep}{0pt}\setlength{\parskip}{0pt}}
\newcommand{\sep}{; }
\providecommand{\Description}[1]{}
\newenvironment{highlights}{\section*{Highlights}\begin{itemize}}{\end{itemize}}
\newenvironment{keyword}{\par\smallskip\noindent\textbf{Keywords: }}{\par\smallskip}

\begin{document}

\title{A semantic mutation metric for metamorphic relation adequacy in scientific computing programs}
\author{
Meng Li\thanks{Corresponding author. Email: mlemon@usc.edu.cn. ORCID: 0000-0002-1074-1502}
\and Xiaohua Yang\thanks{Email: xiaohua1963@foxmail.com. ORCID: 0000-0002-2977-1787}
\and Jie Liu\thanks{Email: jieliu@usc.edu.cn. ORCID: 0009-0008-1970-8347}
\and Shiyu Yan\thanks{Email: yanshiyu@usc.edu.cn. ORCID: 0000-0001-7626-5185}
}
\date{\small
School of Computing, University of South China, Hengyang, China\\
Hunan Engineering Research Center of Software Evaluation and Testing for Intellectual Equipment, Hengyang, China\\
CNNC Key Laboratory on High Trusted Computing, Hengyang, China}
\maketitle
\begin{highlights}
\item Semantic Mutation Score (SMS) generalizes classical Mutation Score (MS)
\item Semantic operators define effect fibers for metamorphic-relation (MR) adequacy
\item 12 Programs Under Test (PUTs), 60 cells: 5.14\% Abstract Syntax Tree (AST) overlap
\item Hyperparameter, Structural Injection, and Trajectory Flip are unreachable by first-order defaults
\item Real-defect evidence separates kill-rate, semantic alignment, and defect detection
\item A strong/weak metamorphic-relation boundary maps false positives and false negatives
\item An adjoint extension arm shows strong relations detecting real adjoint defects
\end{highlights}

\begin{abstract}
\noindent\textbf{Context.} Metamorphic Testing (MT) helps software engineers test scientific programs without full test oracles, but classical Mutation Score (MS) remains syntactic. It does not say whether a metamorphic-relation set observes domain-semantic effects such as conservation, monotonicity, convergence, trajectory, or fidelity-order violations.
\textbf{Objective.} We introduce Semantic Mutation Score (SMS), a backward-compatible adequacy metric for declared semantic strata. The contribution is a metric and certificate model, not a new LLM mutator or benchmark release.
\textbf{Method.} We define five semantic operator families and instantiate them on 12 Programs Under Test (PUTs) across five meta-patterns. Validation combines an analytical SMS-to-MS degeneration result, a 60-cell empirical audit, an Abstract Syntax Tree (AST)-normalised comparison with default first-order syntactic mutation, a diagnostic root-cause layer, a strong/weak metamorphic-relation boundary study, and a compact real-defect evidence slice. The claims are scoped to single-output scientific-computing kernels.
\textbf{Results.} SMS preserves the classical mutation-score denominator while changing only mutant generation, equivalence, and killing. The semantic-mutant pool has 5.14\% AST-normalised overlap with 1,250 default syntactic mutants; three operator families are absent from the default first-order syntactic pool. The pre-registered large-effect threshold is not met and suspect mass is high, but these failures expose operator applicability, MR-design, and attribution boundaries rather than invalidating the metric. The real-defect evidence slice separates aggregate kill-rate from semantic alignment and defect detection.
\textbf{Conclusion.} SMS gives software-testing researchers a reproducible way to report which declared semantic effects an MR set observes, misses, or kills for non-semantic reasons. Its value is diagnostic and construct-level, not a claim of universal dominance.
\end{abstract}


\section{Introduction}\label{introduction}

\subsection{Software-Engineering Motivation and Scope}\label{motivation-and-scope}

Metamorphic Testing (MT) addresses the test-oracle problem in scientific
computing software: instead of checking outputs against a known-correct
reference, MT checks metamorphic relations (MRs), invariants connecting
outputs of related inputs. While \emph{semantic mutation testing} has
been named since \citet{7a6c280d9584691800e71876c716ea229335abeb} and tooled for C \citep{dan2012smtc}
and for UML state-machine behaviour \citep{derezinska2019uml}, and while
MR-adjacent data-semantic mutation \citep{sun2016mumt,sun2024datamutation,zhu2020datamorphic,chan2024metamorphic}
and recent LLM semantic-invariance MT \citep{0fa97cc4e2a0ffb6a777669b1b541c0372fac0d2} extend the mutation target
to test data, relations, and task-level semantics, the adequacy of an MR
set against \emph{domain-specific code-level faults}, conservation laws,
monotonicity, convergence order, trajectory shape, fidelity ordering,
has lacked a backward-compatible metric tied to the classical Mutation
Score (MS) lineage. \citeauthor{d7c38286734419b52de4262c9802ebdfcf4b9447}'s \citeyearpar{d7c38286734419b52de4262c9802ebdfcf4b9447} MS is defined over syntactic
Abstract Syntax Tree (AST) mutations and does not capture these domain
semantics. Recent Large Language Model (LLM) generated mutants \citep{tip2025llmorpheus,humbatova2021deepcrime} do
produce semantically richer mutants, but do not separate the
contribution of LLM source diversity from the contribution of MR design.

Throughout this paper we use the term \emph{four representative classes
of single-output scientific computing kernels} (numeric, probabilistic,
surrogate, and machine-learning), and avoid the ambiguous
software-engineering term \emph{paradigm}. The scope of our claims is
strictly bounded to single-output float-to-float kernels: each
Program Under Test (PUT) is under 2 KB. We do not claim industrial transfer
for multi-output software in this paper.

A compact real-defect evidence slice then checks the same semantic
distinction developed here: mutation kill-rate, semantic stratum
alignment, and real-defect detection are related but different
constructs. The slice strengthens the semantic-mutant argument at result
level; it is not presented as a reusable benchmark artifact. The slice was
selected only after defect/MR detectability verification and is reported
here at result level. Candidate mining, rejected cases, repository
tooling, and benchmark release are outside this manuscript.

The intended reader is a software-testing researcher or practitioner who
needs to decide whether a set of MRs is adequate for a declared semantic
risk, not a domain scientist seeking a new numerical solver or a tool
user seeking another LLM-mutant generator. The novelty is therefore the
adequacy denominator and certificate structure: SMS asks which admitted
semantic-effect fibers are observed by an MR set, while preserving the
classical mutation-score form. The practical impact is diagnostic rather
than deployment-ready automation: the metric exposes when MR design,
operator applicability, or kill attribution prevents a semantic adequacy
claim from being trusted.

\subsection{Three-layer methodological framework}\label{three-layer-methodological-framework}

We organise the contribution as a three-layer framework around
domain-semantic mutation operators.

\begin{itemize}
\tightlist
\item
  \textbf{Layer 1, Definitional (Section~\ref{necessary-conditions-for-semantic-mutation-layer-1}).} A mutation is \emph{semantic}
  if it satisfies at least one of three necessary conditions: (a) it
  crosses a function-call or module-import boundary, (b) it depends on
  domain knowledge for legality, or (c) it changes the algorithmic
  class. The five meta-operator classes, Conservation Erosion (CE, also
  written mut\_C), Operator Substitution (OS, mut\_M), Hyperparameter
  (HP, mut\_G), Trajectory Flip (TF, mut\_T), and Structural Injection
  (SI, mut\_F), specialise these conditions across the four PUT classes.
\item
  \textbf{Layer 2, Operational (Section~\ref{equivalence-judgement-e1-e2-layer-2}, Section~\ref{mutant-pool-prescreen-and-equivalence-detection}).} An equivalence judgement
  E1 $\wedge$ E2, where E1 is Automated Verification Pipeline (AVP) coherence
  and E2 is output equivalence on K\_eq = 1000 samples, gives the
  conservative instantiation of the Layer-1 conditions. The trade-off
  against E1-alone and E2-alone variants is in Appendix A.3.
\item
  \textbf{Layer 3, Applied (Section~\ref{p2-vs-syntactic-mutants-12-put-empirical-layer-3}).} AST-normalised empirical
  traceability across all 12 PUTs: comparing 292 v4 mutants against
  1,250 cosmic-ray syntactic mutants, we show empirically that the v4
  pool is not a subset of the syntactic-mutant pool.
\end{itemize}

The 60-cell empirical audit reported in Section~\ref{statistical-analysis-and-empirical-results} stress-tests this
backbone. SMS is positioned as a strict generalisation of classical MS:
under the
degenerate limit $L = L_{\mathrm{equiv}} \wedge L_{\mathrm{killed}} \wedge L_{\mathrm{mut}}$
formalised in Section~\ref{sms-ms-degeneration-formal-statement}, SMS reduces almost everywhere to MS. Layer 1 and
Layer 3 correspond to the two leading research questions, RQ1
(constructibility and backward compatibility) and RQ2 (syntactic
unreachability); the 60-cell demonstration answers RQ3--RQ5.

The empirical evidence has four roles. The 12-PUT audit is the main
stress test of the SMS construction. The AST audit tests whether the
semantic-mutant space is reachable by default first-order syntactic
mutation. The boundary and adjoint arms map when strong MRs do, and do
not, detect semantic effects. The real-defect slice checks whether the
construct distinctions remain visible on verified MR-detectable defects.
These parts should be read in sequence: the theory defines the metric,
the 12-PUT audit stress-tests one empirical instantiation, the boundary
and adjoint arms test the strong-MR duality, and the real-defect slice
checks whether the same construct separation appears outside synthetic
mutants.

\begin{table}[t]
\caption{Evidence-to-claim map for the empirical argument.}
\label{tab:claim-evidence-map}
\centering
\small
\begin{tabular}{p{0.24\linewidth}p{0.35\linewidth}p{0.33\linewidth}}
\toprule
Claim supported & Evidence used & Claim not made \\
\midrule
SMS is backward-compatible with classical MS & Formal SMS-to-MS
degeneration under the syntactic limit & SMS replaces classical mutation
score in ordinary syntactic settings \\
Semantic mutants occupy a different admitted universe & 5.14\%
AST-normalised overlap with default first-order syntactic mutants;
HP, SI, and TF are absent from the default first-order pool & Default
syntactic mutation tools can never generate higher-order semantic
mutants \\
MR alignment is visible but bounded & Aligned mean SMS 0.275 vs
cross mean SMS 0.061; Cliff's \(\delta\) is positive but below the
pre-registered large-effect threshold & The current 12-PUT audit proves
a large universal aligned-vs-cross effect \\
SMS exposes adequacy boundaries & H1--H4 failures, zero-mass cells,
root-cause labels, and strong-boundary false-positive / false-negative
cases & Failed thresholds are treated as empirical success criteria \\
Real-defect behaviour is distinct from aggregate kill-rate & The
result-level real-defect slice separates semantic alignment, aggregate
kill-rate, and defect detection & The manuscript contributes a reusable
real-defect benchmark \\
\bottomrule
\end{tabular}
\normalsize
\end{table}

This table also marks the paper's non-claims. We do not claim that SMS
is universally superior to classical MS, that pattern-derived MRs
dominate all generic relations, that higher-order mutation has been
refuted, that the real-defect slice is a benchmark contribution, or that
the reported SMS values define industrial acceptance thresholds.

\begin{figure}
\centering
\includegraphics[width=0.85\linewidth,height=\textheight,keepaspectratio,alt={Three-layer methodological framework: Layer 1 (definitional necessary conditions), Layer 2 (operational E\_1 \textbackslash wedge E\_2 judgement), Layer 3 (applied AST-normalised traceability across 12 PUTs).}]{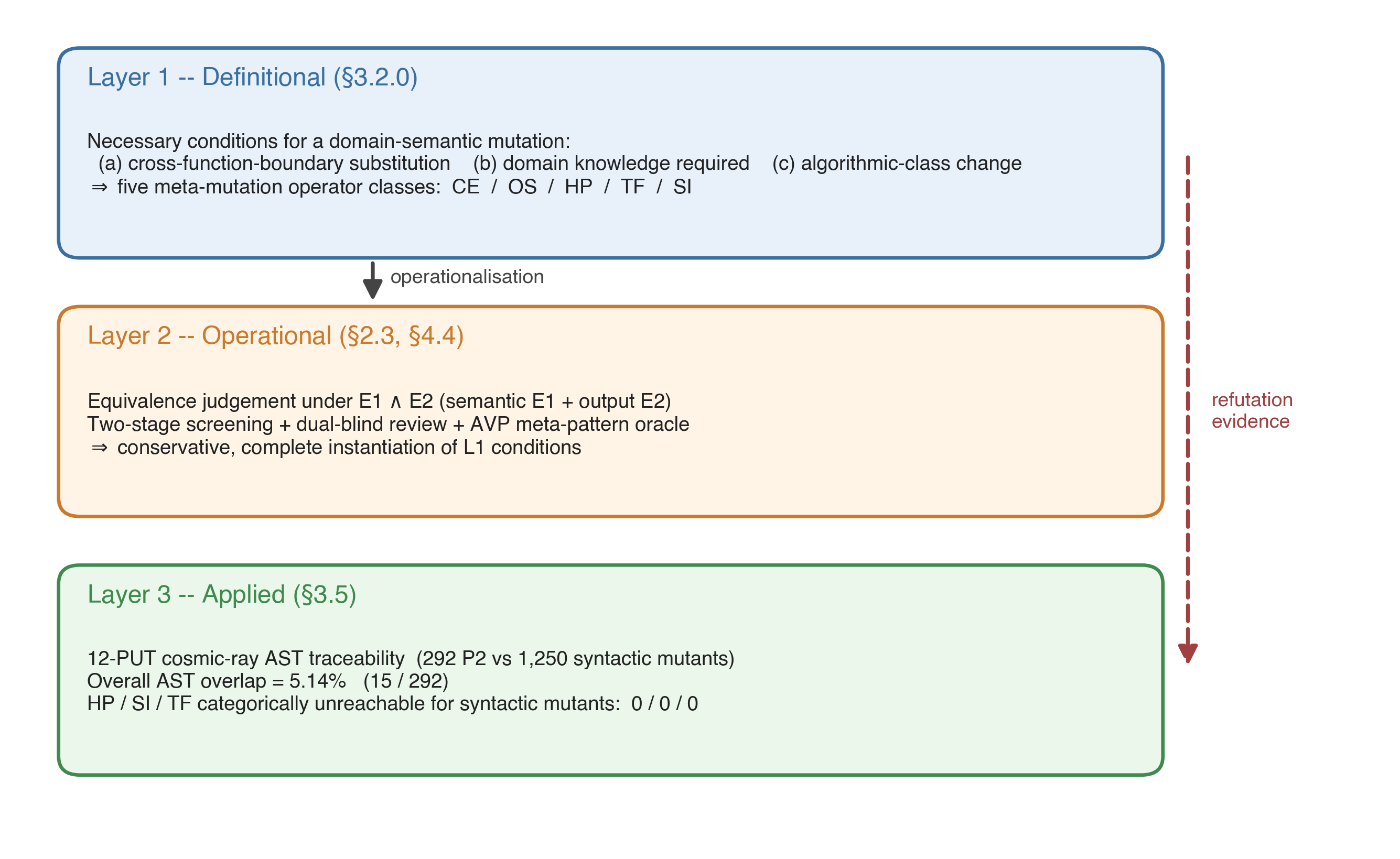}
\caption{Three-layer methodological framework: Layer 1 (definitional
necessary conditions), Layer 2 (operational \(E_1 \wedge E_2\)
judgement), Layer 3 (applied AST-normalised traceability across 12
PUTs).}\label{fig:framework}
\end{figure}

\section{Related Work}\label{related-work-and-roadmap}

Seven lines of prior work bracket the contribution of this paper.

\textbf{(a) Classical mutation testing.} \citet{d7c38286734419b52de4262c9802ebdfcf4b9447} and
\citet{d461ab9482b7fb5eadd7e6cd2c6dffced8ede8b8} survey syntactic mutation. The Coupling Effect
Hypothesis (CPH), that tests detecting simple faults also detect complex
faults, underpins the use of MS as a fault proxy \citep{91b1f655c6e42c02eed722b213e3c03df2e2d75a,andrews2005mutation,just2014mutants}. We
argue in Section~\ref{p2-vs-syntactic-mutants-12-put-empirical-layer-3} that CPH holds within syntactic mutation but does not
automatically extend across the syntactic-versus-domain-semantic
boundary, even when it couples simple syntactic faults to complex
syntactic faults. The Section~\ref{p2-vs-syntactic-mutants-12-put-empirical-layer-3} empirical evidence (5.14\% AST overlap on 12
PUTs, with HP, SI, and TF at 0/0/0) is the empirical witness for this
boundary.

\textbf{(b) Higher-order mutation.} \citet{c03829bdd1a6e45113092384bf0fa05a064ca54a} and \citet{kintis2018effective}
study compositions of first-order syntactic mutants. We
explicitly exclude any equivalence claim about Higher-Order Mutation
(HOM) and list HOM equivalence as a residual threat (Appendix F.2).

\textbf{(c) LLM-generated mutants.} \citet{tip2025llmorpheus}
introduce LLMorpheus, which uses single-LLM JavaScript mutants.
\citet{humbatova2021deepcrime} introduce DeepCrime for
deep-learning real-fault mutation. \citet{dakhel2024llm} extend
LLM-driven mutation testing to test generation on Java PUTs, providing
a recent empirical anchor for the LLM-mutant
lineage. To our knowledge no prior LLM-mutant work isolates \emph{LLM
source diversity} from \emph{MR design} in the contribution to effect
size; the three-stage ablation in Section~\ref{cross-source-v4-protocol-summary} closes this gap. On the
equivalent-mutant problem, \citet{delgadoperez2020equivalent} emphasise
the importance of the \texttt{equiv} term in the MS denominator; we
extend that classical bitwise-equivalence definition to a semantic-class
equivalence E1 $\wedge$ E2 in Section~\ref{equivalence-judgement-e1-e2-layer-2}. \citet{zhang2021classintegration} demonstrate MT
validation in class-integration test ordering, a
non-scientific-computing domain, and the present paper specialises MT to
scientific-computing PUTs and couples it to a domain-semantic mutation
operator framework.

\textbf{(d) Semantic mutation operators across testing communities.}
Semantic mutation as a research line is older than the LLM-mutant
lineage. \citet{7a6c280d9584691800e71876c716ea229335abeb} introduce \emph{semantic
mutation testing} as a path distinct from syntactic mutation; \citet{dan2012smtc}
tool the approach for C; \citet{derezinska2019uml}
apply \emph{semantic mutation operators} to UML state-machine behaviour
by perturbing semantic-variation-point interpretations rather than
syntactic structure. Within MT specifically, \citet{sun2016mumt,sun2024datamutation}
propose \emph{data mutation operators} and the $\mu$MT methodology to drive
MR acquisition through controlled data-semantic perturbation; \citet{zhu2020datamorphic}
formalise \emph{datamorphic testing}, separating
datamorphisms (test-case-to-test-case mappings) from metamorphisms
(test-case-to-Boolean mappings) over data semantics; \citet{chan2024metamorphic}
extend generic data mutation operators to unsupervised
software-defect-prediction validation. Most recently, \citet{0fa97cc4e2a0ffb6a777669b1b541c0372fac0d2}
formalise \emph{semantic invariance} in LLM MT under paraphrase,
fact-reorder, expand/contract, and context-shift transformations. These
works establish that mutation can target data, relation, task, and
behavioural semantics. \textbf{The contribution of this paper is the
first to formalise a domain-specific Semantic Mutation Score (SMS) for
MR adequacy in scientific computing kernels, with a proven
backward-compatible reduction to classical MS (Theorem 9.1) and a
code-level operator framework (Conservation Erosion, Operator
Substitution, Hyperparameter, Trajectory Flip, Structural Injection)
targeting kernel-specific faults, and to delineate the strong boundary
of metamorphic-relation grounding, a strong/weak MR distinction whose
$\varepsilon_{\mathrm{tol}}$-governed duality (Theorem 2, Proposition 2)
is mapped empirically on real and textbook programs spanning the
meta-patterns (Section~\ref{rq4-strong-boundary}), complemented by a cross-library adjoint extension
arm (Section~\ref{adjoint-extension-arm-subjects}, Section~\ref{rq4-adjoint-extension-arm}) that supports the duality forward on third-party
linear operators with real fixed defects.}

\textbf{(e) MR adequacy and coverage.} Recent MT-specific adequacy work
has begun to ask how MR-based tests should be measured rather than
merely generated. The closest literature is best separated by its
\emph{counting object}. \citet{xiang2019compositeMR} construct composite
MRs for GSL scientific functions, but evaluate them with the classical
MuJava denominator of nonequivalent syntactic mutants. \citet{fu2024mtadequacy}
define denominator-bearing MT adequacy over MR obligations and source
inputs; \citet{liu2024mtcoverage} integrate MR coverage into MT adequacy; and
\citet{ba2025metamorphiccoverage} define Metamorphic Coverage over code
elements differentially executed by paired MT runs. These are the
closest denominator-bearing MR/MT coverage comparators, but their
denominators are MR obligations, source-test obligations, or
MR-induced execution coverage rather than semantic mutant classes.
\citet{xie2024mutmodel} quantify MR-set diversity with the MUT model, while
\citet{huang2022mrprioritization} and \citet{zhang2022effectivemrselection} rank or select MRs
using validity, coverage, or mutation-score signals. These latter works
are important boundary surrogates for MR-set quality, but they do not
define an adequacy fraction over an enumerated universe of semantic
fault classes. \citet{302e9d254f138148abce2feda1319f9cb32cd914} provide a useful non-MT analogue
by defining test-suite effectiveness through semantic coverage, showing
that denominator design can move beyond syntax; however, their adequacy
object is a test suite under a specification, not an MR set for
scientific kernels. Thus, prior work measures MR exercise, input
coverage, differential execution, MR diversity, or syntactic-mutant
detection. None makes the adequacy denominator the set of admitted
nonequivalent \emph{domain-semantic code mutants} for scientific
kernels, and none establishes compatibility with the classical MS
denominator. SMS therefore differs not by proposing ``another MR
coverage metric'', but by replacing the denominator against which an MR
set is judged.

\begin{table}[t]
\centering
\footnotesize
\caption{Closest denominator-oriented comparators to SMS. The comparison
is by adequacy object and denominator semantics, not by whether a paper
uses the phrase metamorphic-testing adequacy.}
\label{tab:closest-denominator-comparators}
\begin{tabularx}{\textwidth}{p{0.22\textwidth}p{0.34\textwidth}X}
\toprule
Work & Denominator or counted object & Relation to SMS \\
\midrule
\citep{fu2024mtadequacy,liu2024mtcoverage} & MR obligations, MR coverage, and source-test obligations & Closest MR/MT adequacy comparators, but property/coverage denominators rather than semantic mutant classes \\
\citep{ba2025metamorphiccoverage} & Differentially executed code elements in paired MT runs & Execution-side coverage comparator, not MR-set semantic adequacy \\
\citep{xie2024mutmodel} & MR feature/diversity space & Boundary surrogate for MR-set quality, not an adequacy fraction over faults or semantic classes \\
\citep{huang2022mrprioritization,zhang2022effectivemrselection} & MR validity, coverage, and classical mutation-score signals & Boundary surrogate for cost-effective MR choice, not a denominator-bearing adequacy metric \\
\citep{302e9d254f138148abce2feda1319f9cb32cd914} & Semantic obligations under a specification & Non-MT semantic-coverage analogue; test-suite side rather than MR-set side \\
\bottomrule
\end{tabularx}
\end{table}

\textbf{(f) Property-relative and numerical-specification mutation.}
Recent property-aware mutation work moves the adequacy question closer
to requirements and invariants: \citet{0ed19a2ecb5537dbf1c5f25a818f45a34d418ae8} make mutant
relevance and meaningful killing relative to a tested property, and
\citet{525d0597f3b46f6b8dc6e0ee1e0cd507d23dc22b} generate tests for numerical
specifications through interval-constraint reasoning over floating-point
accuracy and robustness requirements. These works are closer to SMS than
generic AST mutation because they reject property-irrelevant mutants as
an adequacy denominator. They still differ from the present paper in
three respects: the adequacy object here is an MR set rather than a
single formal requirement, the denominator is a five-class
domain-semantic mutant space for scientific kernels, and the metric is
constructed as a backward-compatible generalisation of MS.

\textbf{(g) Semantic foundations and the certificate gap.} Abstract
interpretation supplies the standard machinery for relating concrete
semantics to abstract domains and invariant families \citep{46aba24d3e925f6d8d2efd5b5f6ee0d5ea7558cd},
and for designing program transformations by their effect on an
abstract semantics \citep{72269d5ab33ed790ece6856ba0309ed070d83622}. Approximate program
transformation semantics gives a second, quantitative route: local error
expressions and partial-metric semantics can certify that a
transformation lies within a bounded semantic-effect envelope \citep{11efca8804ed387c9cc3956868f0eb43597c8d10,6d492b783178b02dae163fbfe837857b982deed2}. Refinement- and
UTP-based mutation testing lift mutation to contracts and
specifications \citep{92d99528fe8e70345901cf56aa4d1037f0c3aa96,97a550b469c6151734d1220e37a7985d896191ee}, while
usage-aware differential dynamic logic tracks admissible proof-level
mutations under soundness conditions \citep{7a867d014c3e10e45e4220c5dcf605e8df6ef38d}. These foundations supply much of the machinery needed for
certificate-bearing semantic-effect classes, but they are not MR
adequacy metrics for scientific-computing kernels. Conversely, SMS
provides the MR adequacy denominator and empirical audit, but its
certification layer is empirical and structural rather than
machine-checkable proof-carrying.

\textbf{Closest comparison.} The closest adjacent work is the
Min-MR-Complete formulation \citep{ecf4ccd0f2897516b0205a4b1e35ea81b68aa277}, which uses
the same semantic-mutation family as the fault model for MR-subset
completeness and identifies when class-level reasoning collapses to a
mutant-level set-cover problem. The present paper differs in role: it
defines the SMS denominator, the E1 $\wedge$ E2 admission judgement, the
AST-normalised syntactic-unreachability audit, and the empirical
same-source/cross-source ablation. Min-MR-Complete optimises over an
admitted mutant--draw coverage universe; this paper constructs and
audits that universe. Taken together, the two papers come closest to
semantic-effect-class denominators for MR adequacy, but neither yet
attaches a per-mutant machine-checkable certificate showing that a
syntactic edit belongs to a declared abstract semantic-effect class.

The SMS metric is conceptually complementary to the code-verification
scope of ASME V\&V 20-2009 §3, which targets numerical-solver
correctness. SMS targets MR fault-detection adequacy; we make no
normative compliance claim (Appendix E.3 records this as a long-term
aspiration only).

\section{Problem Formulation and Semantic Mutation Model}\label{research-questions-and-hypotheses}

This section turns the software-engineering problem into auditable
research questions before introducing the metric. The order is deliberate:
scope and novelty are stated as RQ1--RQ2, while RQ3--RQ5 test whether
one empirical instantiation is usable, discriminative, and consistent
enough to support the construct. The formal model then gives reviewers a
small declared vocabulary, the semantic-certificate record, and the
SMS-to-MS degeneration that ties the proposal to the mutation-testing
lineage.

\subsection{Research Questions}\label{research-questions}

The research questions are ordered from the theoretical foundation
(RQ1), through the structural separation that motivates it (RQ2), to the
empirical demonstration on the 60-cell audit (RQ3--RQ5). RQ1 is answered
by construction and proof (Sections~\ref{theory-setting}--\ref{theory-fibers}: Theorem 1 undecidability,
Proposition 1 fiber characterisation, and Theorem 2 duality, together
with the SMS--MS degeneration of Section~\ref{sms-ms-degeneration-formal-statement}); RQ2 by the
AST-normalised structural audit (Section~\ref{p2-vs-syntactic-mutants-12-put-empirical-layer-3}); RQ3--RQ5 by the pre-registered
statistical pipeline (Section~\ref{statistical-analysis-and-empirical-results}). This ordering mirrors the three-layer
framework of Section~\ref{three-layer-methodological-framework}: Layer 1 (definitional) carries RQ1, Layer 3 (applied)
carries RQ2.

\begin{itemize}
\tightlist
\item
  \textbf{RQ1 (Theory of semantic mutation: constructibility,
  undecidability, duality, backward compatibility).} Can semantic
  mutation be formalised as a fiber decomposition of a single effect map
  over admitted nonequivalent \emph{domain-semantic code mutants} for
  scientific-computing kernels, such that (a) stratum membership is
  undecidable and therefore certificate-based (Theorem 1), (b) the
  induced adequacy metric SMS degenerates almost everywhere to the
  classical Mutation Score (Proposition 1; Theorem 9.1), and (c)
  strong metamorphic relations detect only the semantically active
  mutants within a tolerance-bounded regime, while outside it the duality
  degrades along an $\varepsilon_{\mathrm{tol}}$-governed strong boundary
  into false positives (weak MRs, where a correct discrete program
  violates the invariant) and false negatives (surviving non-equivalent
  mutants under coincidental satisfaction) (Theorem 2 and Proposition 2;
  the boundary is mapped empirically under RQ4)?
\item
  \textbf{RQ2 (Syntactic unreachability).} To what extent is the
  domain-semantic mutant space induced by the five operators reachable
  by default-configuration first-order syntactic mutation tools at the
  AST-normalised level?
\item
  \textbf{RQ3 (Instantiability).} Do the five semantic mutation
  operators instantiate a usable nonequivalent semantic-mutant
  denominator across the 60 PUT-MP-operator cells, as reflected by
  inst\_rate, equiv\_rate, C1\_share, and survive\_rate?
\item
  \textbf{RQ4 (Discrimination, LLM-source decoupling, and the strong
  boundary).} Does SMS discriminate operator-MP aligned slices (j=k) from
  cross slices (j$\neq$k); does cross-source LLM pooling under an
  identical prompt move the aligned-versus-cross effect size, or only the
  kill-set quality; and where does the strong boundary of Proposition 2
  lie empirically, on which real and textbook programs does a strong
  MR kill the genuine fault, and where does the duality fail as a weak-MR
  false positive or as a surviving non-equivalent mutant?
\item
  \textbf{RQ5 (Cross-class consistency).} Does the aligned-versus-cross
  SMS pattern generalise across four representative classes of
  single-output scientific-computing kernels? (The relationship between
  SMS and the coverage-style surrogate Pattern Coverage is reported
  descriptively at n = 12 as a hypothesis-generating observation, not a
  confirmatory test.)
\end{itemize}

\subsection{Hypotheses}\label{hypotheses}

The empirical questions RQ3--RQ5 carry pre-registered hypotheses; RQ1 is
answered by Theorem 9.1 and RQ2 by the Section~\ref{p2-vs-syntactic-mutants-12-put-empirical-layer-3} structural audit, neither of
which is a statistical hypothesis.

\begin{itemize}
\tightlist
\item
  \textbf{H1 (RQ3).} At least 4 of the 5 operators produce $\geq$ 5 non-equivalent
  mutants on at least 9 of the 12 PUTs.
\item
  \textbf{H2 (RQ4; pre-registered, exploratory).} The aligned-SMS to
  cross-SMS odds ratio is $\geq$ 3.0, and Cliff's $\delta$ is $\geq$ 0.474.
  This large-effect threshold is an underpowered exploratory target
  (point-estimate verdict and stipulated-alternative power in Section~\ref{rq2-stipulated-alternative-power}), not
  a confirmatory claim of the paper.
\item
  \textbf{H3 (RQ5).} A within-class sign test gives 4/4 across the 4 classes,
  and CV($\Delta$SMS) \textless{} 0.5.
\item
  \textbf{H4 (RQ4).} Mean suspect\_share is $\leq$ 0.20 across the 60 cells.
\end{itemize}

The 12 PUTs are deliberately \emph{Numerical Recipes}--style compact
kernels: they provide a verifiable minimum working example for the
three-layer backbone, while industrial-scale transfer is left to future
work under the representativeness threat.


\subsection{Formal Model and Semantic Mutation Score}\label{notation-and-equivalence-judgement}

\subsubsection{Vocabulary and Notation}\label{vocabulary-inheritance}

We use a small vocabulary throughout the method. The terms are declared
here before they are used in the SMS construction.

\textbf{Program Under Test (PUT).} A PUT is the scientific-computing
kernel being assessed. We write the \(i\)-th PUT as \(S_i\).

\textbf{Metamorphic Relation (MR).} An MR is a property that relates the
outputs of related inputs. The MR family used for PUT \(S_i\) and
meta-pattern \(k\) is written \(\mathrm{MR}_{i,k}\).

\textbf{Meta-pattern (MP).} An MP is a semantic stratum of MR behaviour,
such as conservation, monotonicity, convergence, trajectory structure, or
fidelity order. The index \(k\) denotes the MP being tested.

\textbf{Semantic operator.} A semantic operator is a mutation family
\(\mathrm{mut}_j\) designed to perturb a declared MP. The operator index
\(j\) identifies the intended semantic effect.

\textbf{Semantic mutant.} A semantic mutant \(s'\) is a finite syntactic
edit of \(S_i\) that is admitted by semantic operator \(\mathrm{mut}_j\)
and is certified against the declared MR family. Formally,
\[
  s' \in \mathrm{mut}_j(S_i).
\]
The word semantic therefore names the certification target, not a second
editing mechanism.

\textbf{Semantic-effect fiber.} For an MR family \(\mathrm{MR}_{i,k}\),
the semantic-effect fiber is the subset of admitted semantic mutants
whose declared effect is observable by that MR family:
\[
  \mathcal{F}(\mathrm{MR}_{i,k}) =
  \{s' \in \mathrm{mut}_j(S_i): \mathrm{killed}(s', \mathrm{MR}_{i,k})\}.
\]
Different MR families can observe different fibers. SMS therefore
measures adequacy relative to a declared stratum; it is not a total order
over all possible MR families.

\textbf{Equivalent, killed, and surviving mutants.} A mutant is
equivalent for \((i,k,j)\) when the equivalence judgement of Section~\ref{equivalence-judgement-e1-e2-layer-2} admits
it into \(\mathrm{equiv}_{i,k,j}\). A non-equivalent mutant is killed when
at least one MR in \(\mathrm{MR}_{i,k}\) passes on \(S_i\) and fails on
\(s'\). A non-equivalent mutant that is not killed is surviving. Thus
the generated mutant set decomposes as:

\begin{align*}
\mathrm{mut}_j(S_i) &= \mathrm{equiv}_{i,k,j} \cup \mathrm{killed}_{i,k,j} \cup \mathrm{survive}_{i,k,j} \quad \text{(disjoint)} \\
\mathrm{SMS}_{i,k,j} &:= \frac{|\mathrm{killed}_{i,k,j}|}{|\mathrm{mut}_j(S_i)| - |\mathrm{equiv}_{i,k,j}|} \in [0,\,1]
\end{align*}

\textbf{Semantic Mutation Score (SMS).} SMS preserves the classical
Mutation Score (MS) ratio while replacing the internal definitions of
mutant generation, equivalence, and killing with MR-relative semantic
definitions. The classical concepts of mutation operator, mutant,
equivalent mutant, killed mutant, surviving mutant, and MS are inherited
from \citet{d7c38286734419b52de4262c9802ebdfcf4b9447} and
\citet{ammann2008introduction}.

The complete notation table is in Appendix A.1; index sets and tolerance
parameters ($\varepsilon_{\mathrm{eq}}$, $\varepsilon_{\mathrm{AVP}}^k$,
$K_{\mathrm{eq}} = 1000$, $N = 20$) follow that table.

\subsubsection{Semantic-Operator Signatures}\label{operator-signatures}

Operator family and alignment:

\begin{align*}
\mathrm{MUT} &= \{\mathrm{mut}_C,\, \mathrm{mut}_M,\, \mathrm{mut}_G,\, \mathrm{mut}_T,\, \mathrm{mut}_F\} \quad \text{(open, extensible)} \\
\mathrm{mut}_j &: \text{Programs} \to 2^{\text{Programs}} \\
\mathrm{align}(j) &= j \quad \text{(design choice, not theorem)}
\end{align*}

\begin{table}[t]
\caption{Operator signatures.}\label{tab:p2-01}
\centering
\begin{tabular}{@{}lll@{}}
\toprule
Operator & Failure semantics & Aligned MP \\
\midrule
mut\_C & Conservation-breaking & MP\_1 Conservation \\
mut\_M & Monotonicity-breaking & MP\_2 Monotonicity \\
mut\_G & Convergence-breaking & MP\_3 Convergence \\
mut\_T & Trajectory-distorting & MP\_4 Trajectory \\
mut\_F & Fidelity-order-breaking & MP\_5 Partial-order \\
\bottomrule
\end{tabular}
\end{table}

Per-PUT specialisation tables (mut\_C/M/G/T/F across 12 PUTs and 4
classes) are deferred to \textbf{Appendix B.2}.

\subsubsection{Equivalence Judgement}\label{equivalence-judgement-e1-e2-layer-2}

The equivalence judgement is the executable instantiation of the Layer-1
necessary conditions (Section~\ref{necessary-conditions-for-semantic-mutation-layer-1}). For each candidate mutant
\texttt{s\textquotesingle{}}:

\begin{itemize}
\tightlist
\item
  \textbf{E1 (AVP-coherent):} $\forall$ mr $\in \mathrm{MR}_{i,k}$: $\mathrm{AVP}(S_i, mr)$ =
  $\mathrm{AVP}(s', mr)$.
\item
  \textbf{E2 (Output-equivalent):} $\forall$ x in $K_{\mathrm{eq}}=1000$ samples
  $\sim D_S$: $\|S_i(x) - s'(x)\| \leq \varepsilon_{\mathrm{eq}}$.
\end{itemize}

E1 $\wedge$ E2 is the conservative complete instantiation: false-equiv (a
truly non-equivalent mutant admitted as equivalent) requires \emph{both}
AVP coherence and numerical agreement on K\_eq samples to hold despite the
non-equivalence (both mis-pass simultaneously), which is rare;
false-non-equiv requires only one of E1, E2 to fail for a truly equivalent
mutant, biasing SMS slightly high. The trade-off table against
E1-alone and E2-alone is in Appendix A.3. Under the Section~\ref{sms-ms-degeneration-formal-statement} degenerate
limit \texttt{L}, E1 $\wedge$ E2 reduces almost everywhere to classical bitwise
equivalence (Lemma 9.1).

\paragraph{Semantic certificate record.}
For each admitted mutant, the empirical certificate is a structured
record rather than a proof-carrying object. The record contains the PUT
identifier, operator identifier, intended meta-pattern stratum, edit
class, targeted semantic invariant, E1 AVP-coherence result, E2 sampling
budget and tolerance, killed/survived vector over the relevant MRs, LRCA
labels for killed outcomes, generation source and reviewer status, and a
trace path to the reproduced artifact. A mutant enters the SMS
denominator only after the record passes the Layer-1 semantic-condition
check and the E1 $\wedge$ E2 equivalence judgement. LRCA labels remain
diagnostic annotations on the certificate; they do not filter the
denominator or redefine SMS.

The killed determination preserves the classical OR-aggregation:

\[
\begin{aligned}
\mathrm{killed}(s', \mathrm{MR}_{i,k}) \iff{} & \exists\,mr \in \mathrm{MR}_{i,k}: \\
& \mathrm{AVP}(S_i, mr) = \text{pass} \wedge
  \mathrm{AVP}(s', mr) = \text{fail}.
\end{aligned}
\]

\subsubsection{Root-Cause Diagnostic Layer}\label{lrca-engineering-attribution-layer-descriptive-not-in-sms}

Likely Root-Cause Analysis (LRCA) annotates every killed mutant with one
of five diagnostic labels: C1 likely semantic failure, C2
numerical-tolerance perturbation, C3 out-of-distribution (OOD) trip, C4
statistical-assumption violation, and C5 mutator artefact. The diagnostic
procedure is a three-layer decision tree.
Layer 1 checks tolerance robustness over N = 20 repeats with a
fail-ratio cutoff of 0.80. Layer 2 triages OOD on the surrogate and
machine-learning classes. Layer 3 checks statistical-assumption
baselines on the probabilistic and machine-learning classes using
Wilcoxon and Dynamic Time Warping. A final pass rechecks for mutator
artefacts. Multiple labels are resolved by priority C5 \textgreater{} C4
\textgreater{} C3 \textgreater{} C2 \textgreater{} C1.

The output quantities are \texttt{C1\_share} (share of C1 among killed
mutants) and \texttt{suspect\_share\ :=\ 1\ −\ C1\_share}. \textbf{LRCA
does not modify the SMS formula}: the killed set is never filtered by
suspect status. LRCA is a diagnostic annotation that indicates
whether a given kill is more consistent with a semantic-failure
detection, tolerance sensitivity, OOD behaviour, statistical-assumption
failure, or mutator artefact. Appendix A.2 gives the full decision tree, the 9-grid
threshold calibration, and the engineering rationale for the priority
ordering.

\subsubsection{Backward-compatibility declaration}\label{backward-compatibility-declaration}

The seven core concepts (PUT, mutation operator, mutant, equivalent
mutant, killed mutant, surviving mutant, mutation score) and the SMS
formula
\texttt{\textbar{}killed\textbar{}\ /\ (\textbar{}mut\textbar{}\ −\ \textbar{}equiv\textbar{})}
are aligned item-for-item with \citet{d7c38286734419b52de4262c9802ebdfcf4b9447}. We extend only the
internal definitions: \texttt{mut} shifts from syntactic to
domain-semantic operators, \texttt{equiv} shifts from bitwise
behavioural equivalence to the semantic-class equivalence E1 $\wedge$ E2, and
\texttt{killed} shifts from an equality oracle to a meta-pattern AVP. We
introduce no new formula terms and no new state classifications.

Under the degenerate limit $L$ defined in Section~\ref{sms-ms-degeneration-formal-statement}, SMS reduces
almost everywhere to classical syntactic MS, equivalent mutants
degenerate to classical behavioural equivalence, killed mutants
degenerate to bitwise difference detection, and the LRCA layer
trivialises, C2 through C5 cannot fire when L1 $\wedge$ L2 $\wedge$ L3 hold. Any
SMS-based empirical conclusion is therefore structurally consistent with
the existing mutation-testing literature in the classical
syntactic-mutation regime, so there is no metric-level semantic
fragmentation. The skeleton of eleven concepts, the three-state
decomposition, the SMS formula, and the AVP interface are fixed; the
contents of MUT, MP, C, and $\mathrm{cls}$ remain open to future
extension.

\subsubsection{Degeneration to Classical Mutation Score}\label{sms-ms-degeneration-formal-statement}

The Section~\ref{backward-compatibility-declaration} backward-compatibility claim is formalised as a degeneration
theorem. The full notation cross-reference is \textbf{Appendix G.1}; the
joint conditions and lemma proofs are in \textbf{Appendix G.2-G.4}. We
state only the main theorem and corollary here. Theorem and lemma labels
(Theorem 9.1, Corollary 9.1, Lemma 9.1-9.3) are preserved as stable
identifiers for cross-reference with Appendix G.

\textbf{Definition (Degenerate limit).}
$L = L_{\mathrm{equiv}} \wedge L_{\mathrm{killed}} \wedge L_{\mathrm{mut}}$, where each joint
condition is a pair of paired axes acting on one layer of the SMS
formula:

\begin{itemize}
\tightlist
\item
  $L_{\mathrm{equiv}} = L1 \wedge L2$: $\varepsilon_{\mathrm{eq}} \to 0 \wedge K_{\mathrm{eq}} \to \infty$ (controls equiv
  layer; Lemma 9.1).
\item
  $L_{\mathrm{killed}} = L3 \wedge L4$: $\varepsilon_{\mathrm{AVP}} \to 0 \wedge$ MP set = $\{\mathrm{MP}_{\mathrm{eq}}\}$
  (controls killed layer; Lemma 9.2).
\item
  $L_{\mathrm{mut}} = L5 \wedge L6$: $\mathrm{mut}_j$ switches to rule-based syntactic
  operators (Mothra-style, or modern MutMut/PIT-style) $\wedge$ PUT class $\subseteq$ imperative deterministic
  programs (controls mut layer; Lemma 9.3).
\end{itemize}

\textbf{Theorem 9.1 (SMS $\to$ MS degeneration).} In the degenerate limit
$L$, \textbf{almost everywhere} with respect to the input
distribution $D_S$,

\[ \mathrm{SMS}_{i,k,j} \xrightarrow{L} \mathrm{MS}_{i,j} = \frac{|\mathrm{killed}_{i,j}^{\mathrm{classic}}|}{|\mathrm{mut}_j^{\mathrm{syntax}}(S_i)| - |\mathrm{equiv}_{i,j}^{\mathrm{classic}}|} \]

where the right-hand side is the classical Mutation Score of
\citet{d7c38286734419b52de4262c9802ebdfcf4b9447},
$\mathrm{killed}^{\mathrm{classic}}$ is the difference-detection set,
$\mathrm{equiv}^{\mathrm{classic}}$ is the behavioural-equivalence set,
and $\mathrm{mut}^{\mathrm{syntax}}$ is the syntactic-mutant set.

\textbf{Proof sketch.} By Lemma 9.1, equiv\_\{i,k,j\} $\to$
equiv\_\{i,j\}\^{}classic almost everywhere with respect to D\_S; the
measure-zero qualifier accommodates floating-point pathological points
and Not-a-Number propagation. By Lemma 9.2, killed\_\{i,k,j\} $\to$
killed\_\{i,j\}\^{}classic, and the MP index k degenerates because L4
collapses MR\_\{i,k\} to \{MP\_eq\}. By Lemma 9.3, mut\_j(S\_i) $\to$
mut\_j\^{}syntax(S\_i). Substituting into the SMS formula gives
MS\_\{i,j\}. Full lemma proofs are in Appendix G.3 and the full theorem
proof in Appendix G.4.

\textbf{Corollary 9.1 (LRCA trivialisation).} Under \texttt{L}, the
likely-root-cause inventory C = \{C1, \ldots, C5\} degenerates to
\{C1\}, so suspect\_share $\to$ 0 and SMS becomes a single-layer metric
consistent with the engineering attribution structure of
\citet{d7c38286734419b52de4262c9802ebdfcf4b9447}.

\textbf{Empirical consistency.} Theorem 9.1 + Corollary 9.1 jointly
show that the SMS-based empirical conclusions (Section~\ref{statistical-analysis-and-empirical-results} Cliff's delta,
Friedman chi\^{}2, Spearman rho) are structurally consistent with
existing mutation-testing literature in the classical syntactic-mutation
regime, and \textbf{do not} constitute metric-level semantic
fragmentation.


\subsubsection{Semantic-Mutation Theory Setting}\label{theory-setting}

The preceding SMS-construction subsections fix the SMS metric and its degeneration to MS. We now
give the theory the metric instantiates: semantic mutation is
\emph{intensionally semantic, extensionally syntactic}. Every mutation
is mechanically a syntactic act, because semantics changes only through
syntax. What makes a mutant \emph{semantic} is where its defining
condition lives (at the semantic layer, as a specified invariant
violation) and how its membership is established (by certificate,
because the condition is undecidable).

Fix a grammar $G$ and the set $\mathrm{Prog}(G)$ of well-formed programs
with denotational semantics
$\llbracket\cdot\rrbracket : \mathrm{Prog}(G) \to (\mathcal{X} \rightharpoonup \mathcal{Y})$.
Let $\alpha : \mathcal{Y} \to \mathcal{Y}_\alpha$ be the committed
semantic abstraction (the observable that the AVP of Section~\ref{equivalence-judgement-e1-e2-layer-2} reads), and
write $P \equiv_\alpha P'$ when
$\alpha \circ \llbracket P \rrbracket = \alpha \circ \llbracket P' \rrbracket$
pointwise on the admitted input domain $\mathcal{X}_{\mathrm{adm}}$.
This $\equiv_\alpha$ is exactly the equivalent-mutant relation
$\mathrm{equiv}$ of Section~\ref{vocabulary-inheritance}: the E2 component of the E1$\wedge$E2 judgement
(Section~\ref{equivalence-judgement-e1-e2-layer-2}) is its bounded decision procedure on $K_{\mathrm{eq}}$ samples.

The new ingredient is a finite \emph{semantic invariant family}
$I = \{\psi_1, \dots, \psi_5\}$, where each $\psi$ is a predicate over
the observable behaviour $\alpha \circ \llbracket P \rrbracket$. The five
invariants are exactly the properties the meta-patterns of Section~\ref{operator-signatures} target:
$\psi_1$ conservation, $\psi_2$ monotonicity, $\psi_3$ convergence order,
$\psi_4$ trajectory determinism, $\psi_5$ partial-order consistency, so
$\psi_k$ is the invariant whose violation MP$_k$ detects and whose
breakage $\mathrm{mut}_j$ (at $j=k$) targets. We write
$\llbracket P \rrbracket \models \psi$ when $P$'s observable behaviour
satisfies $\psi$.

The family is open (the extensibility of $I$ noted in Section~\ref{theory-fibers}): a sixth
invariant $\psi_6$, \emph{adjoint consistency}
$\langle A x, y\rangle = \langle x, A^{\mathrm{H}} y\rangle$ for
linear-operator kernels, instantiates the same construction. Its MR is
strong (its violation set is closed under $\equiv_\alpha$), so by Theorem
2 it detects only the semantically active mutants of the adjoint code
path. Because adjoint consistency is meaningful only for linear operators
, not for the chaotic, stochastic, and nonlinear machine-learning
kernels that make up most of the 12 PUTs, $\psi_6$ is \emph{not} folded
into the pre-registered 60-cell design (doing so would leave the invariant
undefined on ten of the twelve subjects); it is instead demonstrated as a
separate confirmatory \emph{adjoint extension arm} on dedicated
third-party linear-operator subjects (Section~\ref{adjoint-extension-arm-subjects}, Section~\ref{rq4-adjoint-extension-arm}), leaving the
pre-registered hypotheses H1--H4 untouched.

\subsubsection{Semantic-Mutant Conditions}\label{theory-s1s5}

The following certified definition is the formal counterpart of the
informal Layer-1 necessary conditions of Section~\ref{necessary-conditions-for-semantic-mutation-layer-1}.

\textbf{Definition (semantic mutant).} Given $(P, \psi)$ with
$\llbracket P \rrbracket \models \psi$, a program $P'$ is a
\emph{semantic mutant of $P$ at stratum $\psi$} iff: (S1)
\emph{realizability}, $P' = e(P)$ for a finite AST edit $e$; (S2)
\emph{baseline}, $\llbracket P \rrbracket \models \psi$; (S3)
\emph{violation with witness},
$\llbracket P' \rrbracket \not\models \psi$, witnessed by some
$x \in \mathcal{X}_{\mathrm{adm}}$ observable through $\alpha$; (S4)
\emph{latency}, $P'$ compiles, terminates on $\mathcal{X}_{\mathrm{adm}}$,
and is type-correct, so the defect is observable only at the semantic
layer, never by a crash oracle; (S5) \emph{stratum purity} (required
where stratum labels feed downstream),
$\llbracket P' \rrbracket \models \psi'$ for all
$\psi' \in I \setminus \{\psi\}$.

A semantic mutation class is therefore defined by which invariant must
break and how (S2--S5) and realized by whichever syntactic edits achieve
it (S1). The operators $\mathrm{mut}_C, \dots, \mathrm{mut}_F$ of Section~\ref{operator-signatures}
are the constructive templates for strata $\psi_1, \dots, \psi_5$.

\textbf{Definition (latency window).} For a violation-magnitude
parameter $\varepsilon$ of an edit template, the \emph{latency window}
is the interval $(\varepsilon_{\mathrm{tol}}, \varepsilon_{\mathrm{crash}})$:
above the tolerance at which the $\psi$-checker witnesses the violation
(S3), below the threshold at which the container crashes (S4). An empty
window certifies non-realizability of the template at that site.

\subsubsection{Effect map, fibers, and three theorems}\label{theory-fibers}

\textbf{Definition (effect map and fibers).} For fixed $P$, the
\emph{effect map} $\sigma$ sends an applicable edit $e$ to the
semantic-effect class of $P_e = e(P)$, valued in
\[
\begin{aligned}
\{\,&\equiv_\alpha,\ \text{ill-formed},\
      \psi_1\text{-viol}, \dots, \psi_5\text{-viol},\\
   &\text{active-off-taxonomy}\,\},
\end{aligned}
\]
where active-off-taxonomy means $P_e \not\equiv_\alpha P$ yet no
$\psi \in I$ flips from satisfied to violated. The \emph{fiber} of a
class is its $\sigma$-preimage. $P_e$ is \emph{semantically active} iff
$P_e \not\equiv_\alpha P$; the strict syntactic-only mutants are exactly
the inactive fiber $\sigma^{-1}(\equiv_\alpha)$.

\textbf{Theorem 1 (undecidability of semantic-mutant membership).} For
Turing-complete $\mathrm{Prog}(G)$ and non-trivial $\psi$, deciding
whether an edit $e$ is a semantic mutant at stratum $\psi$
(conditions S2--S4) is undecidable. \emph{Proof sketch.} S3 requires
deciding $\llbracket P_e \rrbracket \not\models \psi$, a non-trivial
semantic property, undecidable by Rice's theorem; the termination clause
of S4 is independently undecidable by reduction from the halting
problem. Semantic-mutant status therefore cannot be read off the edit;
it can only be certified on bounded fragments with recorded budgets,
which is the formal reason the admission judgement (Section~\ref{equivalence-judgement-e1-e2-layer-2}, Section~\ref{mutant-pool-prescreen-and-equivalence-detection}) is
gate-shaped rather than a decision procedure.

\textbf{Proposition 1 (fiber characterization of classical practice).}
(i) Every semantic mutant arises from some syntactic edit; the
stratum-$\psi$ class equals $\sigma^{-1}(\psi\text{-viol})$ intersected
with S4. (ii) The classical equivalent-mutant problem is exactly the
degenerate fiber $\sigma^{-1}(\equiv_\alpha)$ of unconstrained syntactic
sampling. (iii) Classical mutation testing and semantic mutation are the
same effect map under two sampling policies: sample the edit domain
blindly, versus sample a specified non-degenerate fiber. \emph{Proof.}
(i) holds because semantics changes only through syntax; (ii) and (iii)
follow by unfolding the definitions. The SMS-to-MS degeneration of
Theorem 9.1 (Section~\ref{sms-ms-degeneration-formal-statement}) is the metric-level shadow of (ii): when the admitted
fiber collapses to the syntactic default, the SMS denominator collapses
to the MS denominator.

\textbf{Theorem 2 (duality with strong MRs).} Let $r$ be a
metamorphic relation whose violation set is closed under $\equiv_\alpha$
(a \emph{strong} MR). Then $r$ detects no semantically inactive
mutant, and conversely any mutant detected by a strong MR is
semantically active. Hence over a strong MR universe the kill matrix
is supported exactly on the semantically active fibers, and inactive
(syntactic-only) mutants are structural noise for strong MT evaluation
rather than hard cases. \emph{Proof.} If $P_e \equiv_\alpha P$ and $r$'s
violation predicate factors through $\alpha$, then $r$ flags $P_e$ iff
it flags $P$; MR validity says $r$ does not flag $P$, so it does not
flag $P_e$. The converse is the contrapositive. $\square$

The duality identifies $\alpha$ as the single hinge connecting MR-side
quality (strength) and mutant-side quality (activity). Classification
is relative to $(I, \alpha, \text{budget})$: extending $I$ can only move
a mutant from active-off-taxonomy to $\psi$-stratified, never the
reverse, and bounded $\equiv_\alpha$ verdicts are search conclusions,
not proofs, so every certificate carries its $(I, \alpha, \text{budget})$
stamp. The full reduction for Theorem 1 follows Rice's theorem together
with the standard halting reduction for the termination clause; the
proof of Theorem 2 is the $\equiv_\alpha$-closure argument above.

\textbf{Definition (strong MR, weak MR, strong boundary).} Theorem 2
reads off strength algebraically, assuming the invariant $\psi$ behind
$r$ is satisfied by the correct program so that closure under
$\equiv_\alpha$ makes a violation diagnostic. For a discrete program $P$
approximating a continuous operator this assumption is \emph{contingent}:
the correct $P$ reproduces the operator's algebraic invariant only up to
a discretisation residual, so $\equiv_\alpha$ between $P$ and the ideal
solution holds only within the tolerance $\varepsilon_{\mathrm{tol}}$ of
the latency-window definition. Fixing $\varepsilon_{\mathrm{tol}}$, call
$r$ \emph{strong} on $P$ when the correct $P$ satisfies $r$ within
$\varepsilon_{\mathrm{tol}}$ (its residual stays below
$\varepsilon_{\mathrm{tol}}$ under refinement), so that any violation
beyond $\varepsilon_{\mathrm{tol}}$ certifies a semantically active
mutant and Theorem 2 holds operationally; call $r$ \emph{weak} on $P$
when the correct $P$'s residual exceeds $\varepsilon_{\mathrm{tol}}$, so
that the correct program is itself flagged because the discretisation,
not a fault, breaks the invariant. The \emph{strong boundary} is the
$(\varepsilon_{\mathrm{tol}}, \text{resolution})$ locus separating the
two regimes; tightening $\varepsilon_{\mathrm{tol}}$ or coarsening the
discretisation carries a strong MR across it into the weak regime. A weak
MR is exactly a non-strong (non-grounded) MR in the effective, discrete
sense: its violation set is not closed under the program's
$\varepsilon_{\mathrm{tol}}$-level $\equiv_\alpha$.

\textbf{Proposition 2 (the two boundary failures and their
$\varepsilon_{\mathrm{tol}}$ coupling).} Relative to
$(\varepsilon_{\mathrm{tol}}, P)$ the duality of Theorem 2 has two
failure modes. (i) \emph{False positive (weak MR).} When $r$ is weak the
correct $P$ violates $r$ and a non-fault is reported; the
rotational-symmetry invariant of a neutron-transport operator under a
method-of-characteristics discretisation, satisfied only at special
angles, is the canonical instance. (ii) \emph{False negative (surviving
non-equivalent mutant).} A semantically active mutant
$P_e \not\equiv_\alpha P$ can leave every invariant in $I$ satisfied
within $\varepsilon_{\mathrm{tol}}$, a \emph{coincidental satisfaction
of the MR}, and so survive the strong MR family; this surviving mutant
is non-equivalent (genuinely $\not\equiv_\alpha P$) and is therefore
distinct from the classical equivalent mutant of Proposition 1(ii), and
when its signature lies below $\varepsilon_{\mathrm{tol}}$ it is a
\emph{tolerance fault}. The two modes are coupled through
$\varepsilon_{\mathrm{tol}}$: tightening it shrinks the false-negative
set (more surviving mutants are killed) but grows the false-positive set
(more correct programs are flagged), and loosening it does the reverse.
The boundary is therefore an $\varepsilon_{\mathrm{tol}}$-governed
\emph{sensitivity--specificity trade-off}, not a defect to be engineered
away. \emph{Proof.} (i) and (ii) instantiate the Definition: weakness
places the correct program's residual above $\varepsilon_{\mathrm{tol}}$,
and coincidental satisfaction places the mutant's signature below it;
monotonicity of both sets in $\varepsilon_{\mathrm{tol}}$ holds because
both membership predicates are threshold comparisons against the same
$\varepsilon_{\mathrm{tol}}$. $\square$

Theorem 2 thus holds within the strong regime, and the
$\varepsilon_{\mathrm{tol}}$-governed strong boundary, where a strong
MR degrades into a weak MR (false positives) or is evaded by a surviving
non-equivalent mutant (false negatives), is mapped empirically on real
and textbook programs in Section~\ref{rq4-strong-boundary}.


\section{Study Design}\label{experimental-subjects-and-operator-framework}

The study design follows three validation paradigms rather than relying
on a single small experiment. First, analytical validation establishes
that SMS is a conservative extension of classical MS. Second, the
controlled 12-PUT, 60-cell audit tests constructibility, syntactic
separation, aligned-versus-cross discrimination, and root-cause
attribution under a reproducible protocol. Third, boundary, adjoint, and
real-defect evidence check whether the construct behaves as expected
outside the main synthetic-mutant grid. The design is intentionally
bounded: compact kernels make semantic certification inspectable, while
industrial transfer is treated as a future validity question rather than
assumed.

\subsection{Program Selection}\label{put-selection-12-puts-4-classes}

{\def\LTcaptype{table} 
\small
\begin{longtable}[]{@{}
  >{\raggedright\arraybackslash}p{(\linewidth - 8\tabcolsep) * \real{0.2000}}
  >{\raggedright\arraybackslash}p{(\linewidth - 8\tabcolsep) * \real{0.2000}}
  >{\raggedright\arraybackslash}p{(\linewidth - 8\tabcolsep) * \real{0.2000}}
  >{\raggedright\arraybackslash}p{(\linewidth - 8\tabcolsep) * \real{0.2000}}
  >{\raggedright\arraybackslash}p{(\linewidth - 8\tabcolsep) * \real{0.2000}}@{}}
\caption{PUT selection.}\label{tab:p2-02}\\
\toprule\noalign{}
\begin{minipage}[b]{\linewidth}\raggedright
Class
\end{minipage} & \begin{minipage}[b]{\linewidth}\raggedright
PUT
\end{minipage} & \begin{minipage}[b]{\linewidth}\raggedright
Name
\end{minipage} & \begin{minipage}[b]{\linewidth}\raggedright
Mathematical structure
\end{minipage} & \begin{minipage}[b]{\linewidth}\raggedright
LOC
\end{minipage} \\
\midrule\noalign{}
\endfirsthead
\endhead
\bottomrule\noalign{}
\endlastfoot
\textbf{A Numeric} & A1 & Lorenz ODE integration & Nonlinear ODE &
\textasciitilde150 \\
& A2 & LU decomposition & Linear algebra & \textasciitilde80 \\
& A3 & FDM 1D heat conduction & Parabolic PDE & \textasciitilde200 \\
\textbf{B Probabilistic} & B1 & Beta-Binomial conjugate & Analytic
posterior & \textasciitilde60 \\
& B2 & MCMC Metropolis-Hastings & Markov chain & \textasciitilde250 \\
& B3 & Monte Carlo integration & Importance sampling &
\textasciitilde100 \\
\textbf{C Surrogate} & C1 & Gaussian Process Regression & Kernel methods
& \textasciitilde300 \\
& C2 & Polynomial Chaos Expansion & Orthogonal basis &
\textasciitilde250 \\
& C3 & Neural-net surrogate & MLP substitution & \textasciitilde400 \\
\textbf{D ML} & D1 & Multi-Layer Perceptron & Backprop. &
\textasciitilde350 \\
& D2 & Support Vector Machine & Convex optimisation &
\textasciitilde200 \\
& D3 & Logistic Regression & Maximum likelihood & \textasciitilde120 \\
\end{longtable}
}

The 12 PUTs are taken from the shared 12-PUT infrastructure \citep{ecf4ccd0f2897516b0205a4b1e35ea81b68aa277} but
independently justified along four dimensions: library-stack coverage
(numpy 2.4.4, scipy 1.17.1, scikit-learn 1.8.0); mathematical-structure
coverage (8 of 12 chapters of \emph{Numerical Recipes}); overlap with
existing mutation-testing benchmarks (DeepCrime, Defects4J, and the
mutmut and cosmic-ray demos); and signature-simplification trade-offs.
The full coverage argument is in Appendix B.1. The
\texttt{program(x:\ float)\ $\to$\ float} signature is a substantive
constraint (Section~\ref{threats-to-validity}) that bounds the upper limit of mutant semantic
complexity; transfer to industrial multi-output PUTs is left to
outside the evaluated scope of this paper.

\subsection{Necessary conditions for semantic mutation}\label{necessary-conditions-for-semantic-mutation-layer-1}

\textbf{Definition (Semantic mutation criteria).} A mutant
\texttt{s\textquotesingle{}\ =\ mut\_j(S\_i)} is a \emph{semantic}
mutation if and only if it satisfies at least one of the following three
conditions.

\begin{enumerate}
\item
  \textbf{Cross-function-boundary replacement.} The AST node operated on
  crosses at least one function-call or module-import boundary. For
  example, \texttt{np.linalg.det(M)\ $\to$\ np.sum(np.diag(M))}.
\item
  \textbf{Carries domain knowledge.} The legality of the mutation
  depends on mathematical, physical, or statistical knowledge of the
  program's domain, not on purely syntactic type preservation. For
  example, changing the Gaussian Process Regression
  \texttt{noise\_level} from 1e-4 to 1e-1.
\item
  \textbf{Changes algorithmic class.} The mutation alters the
  algorithmic class implemented. For example, replacing RK4 with Euler
  changes the integration order.
\end{enumerate}

A mutation that satisfies none of (a) -- (c) is purely \emph{syntactic}:
AST-local, domain-agnostic, and class-preserving. The five meta-operator
classes (CE, OS, HP, TF, SI) are specialisations of (a) -- (c):

{\def\LTcaptype{table} 
\begin{longtable}[]{@{}lllll@{}}
\caption{Necessary conditions for semantic mutation.}\label{tab:p2-03}\\
\toprule\noalign{}
Operator class & (a) & (b) & (c) & Primary condition \\
\midrule\noalign{}
\endfirsthead
\endhead
\bottomrule\noalign{}
\endlastfoot
\textbf{CE} Constant perturbation & $\times$ & $\bigtriangleup$ & $\times$ & partial (b); weakest \\
\textbf{OS} API replacement & \checkmark & \checkmark & $\bigtriangleup$ & (a)+(b) \\
\textbf{HP} Hyperparameter & $\times$ & \checkmark & $\bigtriangleup$ & (b)+partial(c) \\
\textbf{TF} Numerical transform & $\bigtriangleup$ & \checkmark & \checkmark & (b)+(c) \\
\textbf{SI / CF} Structural injection & $\bigtriangleup$ & \checkmark & \checkmark & (b)+(c) \\
\end{longtable}
}

Only \textbf{CE partially satisfies} the necessary conditions (it is a
semantic / syntactic boundary class); OS, HP, TF, SI strongly satisfy at
least one of (a), (b), (c).

\subsection{Metamorphic-Relation Coverage Matrix}\label{cell-instantiation-matrix}

{\def\LTcaptype{table}
\small
\begin{longtable}[]{@{}lccccc@{}}
\caption{Metamorphic-relation coverage matrix: per-PUT coverage density across the five meta-patterns (12 PUTs $\times$ 5 meta-patterns). $\bullet\bullet$ substantial (30 cells), $\bullet$ moderate (24 cells), $\circ$ vacant (6 cells; no MR is exercised).}\label{tab:p2-cells}\\
\toprule\noalign{}
PUT & MP\_1 cons. & MP\_2 mono. & MP\_3 conv. & MP\_4 traj. & MP\_5 p-ord. \\
\midrule\noalign{}
\endfirsthead
\endhead
\bottomrule\noalign{}
\endlastfoot
A1 Lorenz  & $\bullet\bullet$ & $\bullet$ & $\bullet\bullet$ & $\bullet\bullet$ & $\circ$ \\
A2 LU      & $\bullet\bullet$ & $\circ$ & $\bullet$ & $\bullet$ & $\bullet\bullet$ \\
A3 FDM     & $\bullet\bullet$ & $\bullet$ & $\bullet\bullet$ & $\bullet\bullet$ & $\circ$ \\
B1 BetaBin & $\bullet\bullet$ & $\bullet$ & $\circ$ & $\circ$ & $\bullet$ \\
B2 MCMC    & $\bullet$ & $\bullet\bullet$ & $\bullet\bullet$ & $\bullet\bullet$ & $\bullet$ \\
B3 MC      & $\bullet\bullet$ & $\circ$ & $\bullet\bullet$ & $\bullet$ & $\circ$ \\
C1 GPR     & $\bullet$ & $\bullet\bullet$ & $\bullet\bullet$ & $\bullet$ & $\bullet\bullet$ \\
C2 PCE     & $\bullet\bullet$ & $\bullet$ & $\bullet\bullet$ & $\bullet$ & $\bullet\bullet$ \\
C3 NN-Surr & $\bullet$ & $\bullet\bullet$ & $\bullet\bullet$ & $\bullet\bullet$ & $\bullet\bullet$ \\
D1 MLP     & $\bullet\bullet$ & $\bullet\bullet$ & $\bullet$ & $\bullet$ & $\bullet\bullet$ \\
D2 SVM     & $\bullet$ & $\bullet\bullet$ & $\bullet$ & $\circ$ & $\bullet\bullet$ \\
D3 LR      & $\bullet\bullet$ & $\bullet\bullet$ & $\bullet$ & $\circ$ & $\bullet\bullet$ \\
\end{longtable}
}

\noindent\emph{Legend:} $\bullet\bullet$ substantial (30 cells); $\bullet$ moderate (24 cells); $\circ$ vacant (6 cells; no MR exercised).

Each PUT averages 24.3 mutants in the cross-source pool (range
10-30), for 292 unique v4 mutants across the 12 PUTs; each mutant is
evaluated in its relevant cells over the 60-cell design with N=20 AVP
repetitions per cell, and K\_eq = 1000 input samples for E2 evaluation. Detailed
pool counts and per-class operator specialisations are in
\textbf{Appendix B.2}.

The cell density notation distinguishes \textbullet\textbullet substantial (30 cells;
aligned slice + strong off-diagonal cells where MR coverage is dense and
mutant detection is expected), \textbullet moderate (24 cells; cross slice with
non-trivial expected detection rate), and $\circ$ vacant (6 cells;
historically empty cells where no MR is exercised). The 12
aligned (j = k) cells are precisely the on-diagonal cells; the 48 cross
(j $\neq$ k) cells partition into the 24 \textbullet moderate + 18 \textbullet\textbullet off-diagonal
substantive + 6 $\circ$ vacant cells. The \textbullet moderate / \textbullet\textbullet substantive
distinction does not change SMS computation; it tracks expected MR
coverage density per the infrastructure documentation and informs the Section~\ref{decoupling-between-r_sem-and-r_kill} R\_sem /
R\_kill decoupling discussion.

\textbf{Per-class operator specialisations (illustrative).} Each
meta-operator (mut\_C / M / G / T / F = CE / OS / HP / TF / SI in their
dual form) requires PUT-class-specific specialisation:

\begin{itemize}
\tightlist
\item
  \textbf{mut\_C Conservation-breaking:} A1 Lorenz adds $\varepsilon_{\mathrm{drift}}$ to RHS
  (slow Hamiltonian drift); A2 LU decomposition omits the (k+1)-th row
  multiplier; B1 Beta-Bin posterior omits normalisation; C1 GPR
  covariance omits positive-definite diagonal term; D1 MLP backprop
  omits one gradient term.
\item
  \textbf{mut\_M Monotonicity-breaking:} A3 FDM $\Delta$t occasionally
  negative; B2 MCMC acceptance min(1, r) $\to$ min(0.95, r); C2 PCE
  high-order coefficient sort inserts inversion; D2 SVM
  decision-function sign flips near boundary.
\item
  \textbf{mut\_G Convergence-breaking:} A1 Lorenz RK4 $\to$ 1.5-order
  hybrid; A3 FDM 2nd-order difference $\to$ 1st-order; B3 MC doubling sample
  size does not 1/N-reduce variance; C3 NN-Surr training-epoch
  truncation.
\item
  \textbf{mut\_T Trajectory-distorting:} A1 Lorenz state-vector y / z
  swap; B2 MCMC inserts independent-sampling segment; C3 NN-Surr
  training-target slow phase shift; D1 MLP hidden-layer periodic-mask
  activation.
\item
  \textbf{mut\_F Fidelity-order-breaking:} A2 LU partial pivoting
  degrades to no pivoting; C1 GPR length-scale switches to coarse prior;
  C2 PCE high-order term randomly retains low-order; D3 LR
  regularisation occasionally large.
\end{itemize}

The per-class HP / OS / TF substitution rules give similarly
differentiated specialisations across PUT classes (e.g., HP on class a =
tolerance / max\_iter, on class c = GPR \texttt{noise\_level} /
\texttt{length\_scale}, on class d = MLP \texttt{hidden\_dim} / dropout;
OS on class a = numerical-linalg API swap \texttt{det} $\leftrightarrow$
\texttt{sum(diag)}, on class b = sampling-API swap, on class c =
surrogate-class swap GPR $\leftrightarrow$ RBF $\leftrightarrow$ NN). The full PUT-class $\times$ operator
specialisation grid is in Appendix B.2.

\subsection{Primary Meta-Pattern Convention}\label{primary-mp-convention}

The diagonal cells \texttt{j\ =\ k} form the H2-aligned slice, the
off-diagonal cells form the cross slice, and the vacant cells \texttt{$\circ$}
are not formally adjudicated.

The c-class primary MP is held at the pre-registered choice (MP5)
throughout this paper. All H1, H2, H3, and H4 verdicts are rendered
under this configuration on v3 (same-source) and v4 (cross-source under
an identical prompt). An earlier exploratory data-driven primary-MP
shift was considered but withdrawn from the analysis after a permutation
null over fully exchangeable c-class (PUT, MP) cell SMS values showed
the resulting $\delta$ inflation to be statistically indistinguishable from
random reselection of the c-class primary MP; the corresponding
pre-registration of a primary-MP-selection rule on a fresh dataset is
deferred to a follow-up study.

\begin{figure}
\centering
\includegraphics[width=0.8\linewidth,height=\textheight,keepaspectratio,alt={Per-class AST overlap rate between 292 semantic mutants and 1,250 cosmic-ray syntactic mutants across 12 PUTs (overall 5.14\%). HP, SI, TF (54.5\% of the semantic pool) are categorically unreachable (0/72, 0/33, 0/54); CE 7.81\%, OS 11.67\%, CF 33.33\% (CF is a class-specialisation, not in main 5).}]{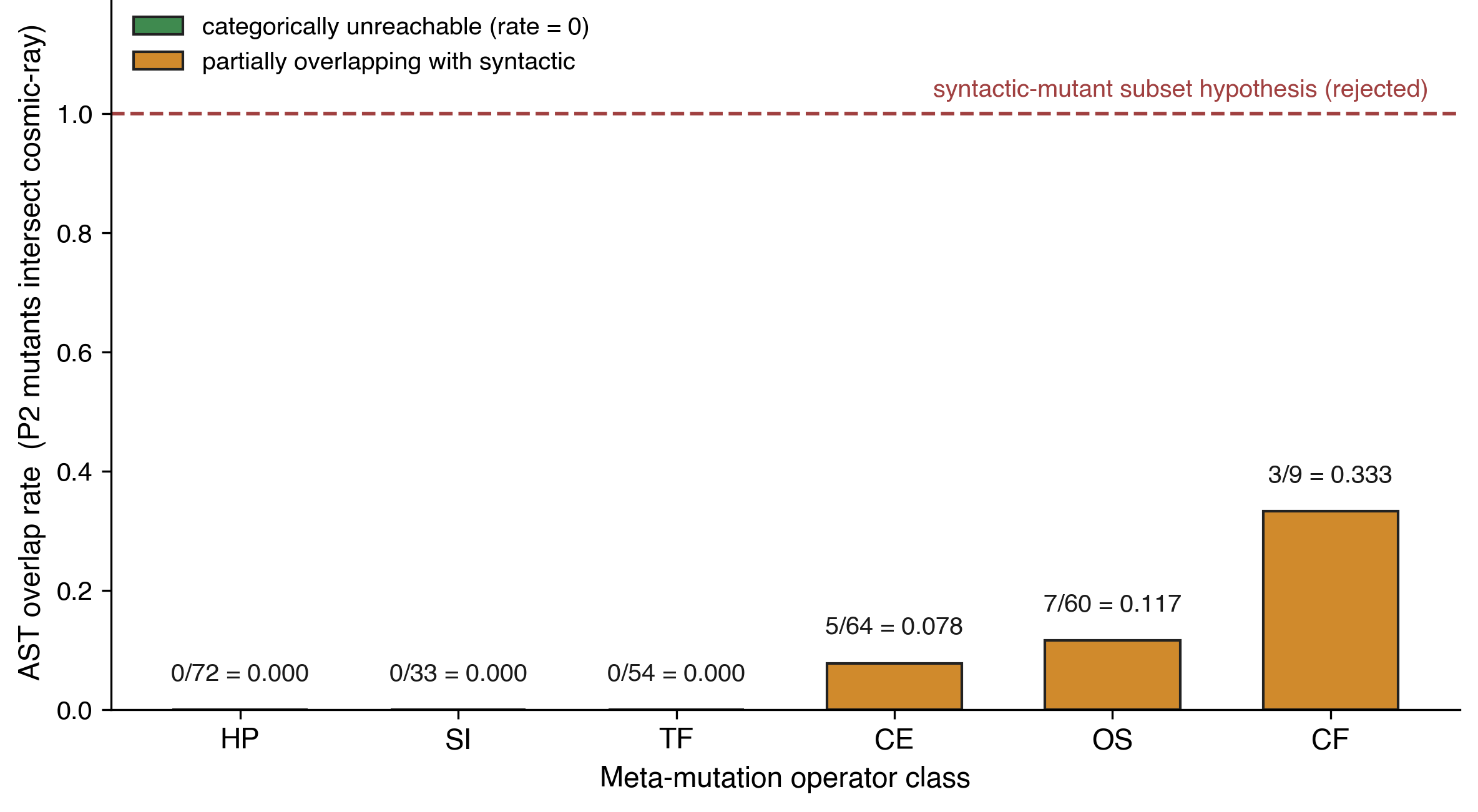}
\caption{Per-class AST overlap rate between 292 semantic mutants and 1,250
cosmic-ray syntactic mutants across 12 PUTs (overall 5.14\%). HP, SI, TF
(54.5\% of the semantic pool) are categorically unreachable (0/72, 0/33, 0/54); CE
7.81\%, OS 11.67\%, CF 33.33\% (CF is a class-specialisation, not in
main 5).}\label{fig:overlap}
\end{figure}

\subsection{Abstract-Syntax-Tree Overlap against Syntactic Mutants}\label{p2-vs-syntactic-mutants-12-put-empirical-layer-3}

\textbf{Experimental design.} For each PUT, semantic cross-source mutants
(\texttt{data/mutants/\$\{PUT\}\_pool\_v4/}, 292 total) are compared at
the AST-normalised level
(\texttt{ast.dump(annotate\_fields=False,\ include\_attributes=False)})
with cosmic-ray default-operator mutants (1,250 total;
\texttt{scripts/p2\_vs\_syntactic\_ast\_diff\_batch.py}). Source:
\texttt{data/results/cosmic\_ray\_12put\_ast\_diff.json}.

\textbf{Aggregate results.}

{\def\LTcaptype{table} 
\begin{longtable}[]{@{}ll@{}}
\caption{Semantic vs syntactic mutants: 12-PUT empirical.}\label{tab:p2-04}\\
\toprule\noalign{}
Metric & Value \\
\midrule\noalign{}
\endfirsthead
\endhead
\bottomrule\noalign{}
\endlastfoot
Total semantic mutants (12 PUTs) & 292 \\
Total cosmic-ray syntactic mutants (12 PUTs) & 1,250 \\
AST-normalised overlap & 15 \\
\textbf{Overall overlap rate} & \textbf{5.14\%} \\
\end{longtable}
}

\textbf{Per-operator-class breakdown (12-PUT aggregate).}

{\def\LTcaptype{table} 
\begin{longtable}[]{@{}lllll@{}}
\caption{Semantic vs syntactic mutants: 12-PUT empirical.}\label{tab:p2-05}\\
\toprule\noalign{}
Class & n\_p2 & n\_overlap & Rate & Interpretation \\
\midrule\noalign{}
\endfirsthead
\endhead
\bottomrule\noalign{}
\endlastfoot
\textbf{HP} & 72 & 0 & \textbf{0.000} & Structurally unreachable \\
\textbf{SI} & 33 & 0 & \textbf{0.000} & Structurally unreachable \\
\textbf{TF} & 54 & 0 & \textbf{0.000} & Structurally unreachable \\
CE & 64 & 5 & 0.078 & Boundary class (Section~\ref{necessary-conditions-for-semantic-mutation-layer-1} partial (b)) \\
OS & 60 & 7 & 0.117 & Partial incidental hits \\
CF & 9 & 3 & 0.333 & b2 only, n=9 \\
\end{longtable}
}

\textbf{Interpretation: refuting the ``post-classification copy''
challenge.}

\begin{itemize}
\tightlist
\item
  HP, SI, and TF (159 of 292 = 54.5\% of v4 mutants) are
  \textbf{unrepresentable under default first-order configurations} of
  cosmic-ray's AST-local operators such as BinOp, Compare, and
  NumberReplacer. The unreachability is structural under first-order
  operators; whether HOM-based compositions \citep{jia2008subtle} (not
  refuted) can close the gap is a residual higher-order mutation threat, addressed
  conceptually in Section~\ref{preventive-defence-framing}.
\item
  The CE class shows 7.81\% incidental overlap, mostly LLM-generated
  half-step perturbations on a2 and b3 such as
  \texttt{\_RHO=28.0\ $\to$\ 27.5}. The remaining 92.19\% of CE mutants are
  AST-disjoint, because the LLMs prefer domain-aware perturbations over
  integer ±1 changes.
\item
  An earlier draft labelled the OS row ``$\times$ tool-inexpressible''. The
  12-PUT data refines this categorical claim to ``$\bigtriangleup$ 88.33\% disjoint,
  11.67\% incidental hits''. The systematic-versus-incidental argument
  in Section~\ref{preventive-defence-framing} (with details in Appendix B.1.5) clarifies why those
  incidental hits are stochastic byproducts rather than systematic
  semantic mutation.
\item
  94.86\% of the v4 mutants cannot be reproduced by cosmic-ray defaults.
  The two pools occupy systematically distinct mutant spaces.
\end{itemize}

A multi-tool cross-comparison was not run because mutpy is incompatible
with Python 3.10+, and mutmut's operator set overlaps strongly with
cosmic-ray's. The DeepSeek, Claude, and GPT contributions to the 15
overlap files are uneven (DeepSeek 11, Claude 4, GPT 0). Appendix B.3
covers the cosmic-ray a1 single-PUT pre-12-PUT pilot, and Appendix F
discusses the LLM-source distributional-shift threat.

\subsection{Higher-Order Mutation Boundary}\label{preventive-defence-framing}

\textbf{Scope of the claim.} The preventive-defence claim below is
conditional on a first-order syntactic baseline; HOM-based syntactic
compositions are an open question and are not refuted by the Section~\ref{p2-vs-syntactic-mutants-12-put-empirical-layer-3}
evidence.

The HP, SI, and TF zero-overlap result, combined with the Section~\ref{necessary-conditions-for-semantic-mutation-layer-1}
necessary-conditions argument, amounts to a \emph{preventive-defence}
argument: semantic mutation operators address a class of fault
hypotheses that lie beyond the reach of default first-order syntactic
configurations (HOM not refuted; see Section~\ref{preventive-defence-framing}). Three points sharpen
this framing.

\textbf{(a) Systematic versus incidental.} Satisfying conditions (a),
(b), or (c) is sufficient for a single semantic mutation, but only when
satisfaction reflects design intent, not stochastic byproduct, does it
constitute a \emph{systematic} semantic mutation method. A syntactic
tool that occasionally hits (a) or (c) with its 12 default operators
does so at a non-zero probability, but the hits are not repeatable and
they carry neither of the two engineering goods we want. First,
designing a semantic mutator like OS \texttt{det\ $\to$\ sum(diag)} requires
knowing that these expressions are equivalent on diagonal matrices but
not on general matrices, a deepening of source-code understanding the
syntactic tool does not exhibit. Second, domain-semantic faults, wrong
physical constants, unit conversions, boundary conditions,
hyperparameter semantics, numerical-method order, are not AST-local; the
syntactic-tool design goals (operator typos, off-by-one errors, negation
flips) do not target them. Systematic semantic mutation therefore
requires (a), (b), or (c) to be design intent. Appendix B.1.5 records
this argument in full.

\textbf{(b) HOM caveat.} HOM \citep{c03829bdd1a6e45113092384bf0fa05a064ca54a,kintis2018effective}
could in principle compose syntactic mutants, for example, an Arithmetic
Operator Replacement combined with a Statement Deletion, to partially
simulate the effects of OS or HP on some PUTs. The Section~\ref{statistical-analysis-and-empirical-results} empirics did not
run a HOM comparison; we list HOM equivalence testing as residual threat
higher-order mutation equivalence (Appendix F.2) and confine our ``tool unreachability'' claim to
\emph{first-order} syntactic tools (mutmut and cosmic-ray default
configurations).

\textbf{Conceptual analysis of HOM reachability.} Higher-order mutation
(HOM; \citealp{jia2008subtle,c03829bdd1a6e45113092384bf0fa05a064ca54a,kintis2018effective}) composes multiple first-order operators on the same
source location. A natural challenge to this paper's preventive-defence
framing is whether HOM compositions of cosmic-ray's
13 default operators or mutmut's 14 default operators can reach the Section~\ref{necessary-conditions-for-semantic-mutation-layer-1}
necessary conditions (a) cross-function-boundary, (b)
domain-knowledge-dependent, or (c) algorithmic-class-change. We address
this conceptually without empirical HOM testing.

For (a): HOM compositions remain over first-order operators that are
AST-local within their target node; chaining AOR $\circ$ Statement-Deletion
within a single function does not introduce a cross-module substitution
like \texttt{det(M)\ $\to$\ sum(np.diag(M))}, which presupposes that
\texttt{np.diag} is in scope and semantically equivalent on diagonal
matrices but not on general matrices. HOM can reach a kindred
destination only if both operators happen to coincide with a meaningful
cross-boundary swap, which is exponentially unlikely under the default
operator menus.

For (b): HOM has no awareness of which constants are load-bearing in the
PUT class. A two-operator chain such as Number-Replacer (1e-4 $\to$ 1e-3) $\circ$
Boolean-Swap on a Gaussian-process kernel does not ``know'' that 1e-4 is
the regularization knob; it modifies the literal numerically without
crossing the domain-semantics boundary. The Hyperparameter operator
class (HP, mut\_G) is by construction parameter-aware and
class-specific, a property HOM compositions do not exhibit by default.

For (c): Algorithmic-class change (e.g.~polynomial $\to$ spline basis;
Lorenz RK4 $\to$ 1.5-order hybrid) requires structural code rewrites at the
level of function bodies, library imports, or iteration schemes. HOM
compositions over default first-order operators are bounded above by
AST-local edits and so cannot reach algorithmic-class change without
being effectively equivalent to a manual rewrite.

We therefore assert that the Section~\ref{p2-vs-syntactic-mutants-12-put-empirical-layer-3} 0/0/0 unreachability for HP / SI / TF,
while strictly empirical only for first-order configurations, is
unlikely to be refuted by default-operator HOM compositions on
combinatorial grounds. A direct empirical HOM falsification (generating
mutmut second-order mutants and re-running the AST overlap analysis)
remains a residual higher-order mutation threat.

A multi-syntactic-tool cross-comparison (mutmut and mutpy) was not run:
mutpy is incompatible with Python 3.10+, and mutmut's operator
set strongly overlaps cosmic-ray's. The full operator-class
cross-reference between cosmic-ray (13 default operators) and mutmut (14
default operators), including which Section~\ref{necessary-conditions-for-semantic-mutation-layer-1} necessary conditions each
operator family reaches, is in Appendix B.6. The aggregate of the two
default sets (\textasciitilde21 distinct first-order AST-local operator
classes) collectively fails all three Section~\ref{necessary-conditions-for-semantic-mutation-layer-1} necessary conditions.

\textbf{(c) Source distributional shift.} The 15 cosmic-ray AST
overlaps are not evenly distributed across the three LLMs (DeepSeek
11/15, Claude 4/15, GPT 0/15; see
\texttt{cosmic\_ray\_12put\_ast\_diff.json}). DeepSeek tends to generate
syntactically simpler mutations. This LLM-source bias is discussed as a
distributional-shift threat
(Appendix F.2) but does \textbf{not} affect the systematic-vs-incidental
argument, which is based on \emph{categorical} AST-locality, not hit
frequency: the HP / SI / TF zero-overlap result is 0/0/0 across all
three sources.

\subsection{Adjoint Extension Subjects}\label{adjoint-extension-arm-subjects}

The adjoint invariant $\psi_6$ (Section~\ref{theory-setting}) is exercised on a separate,
confirmatory set of three third-party linear-operator libraries:
scipy \texttt{LinearOperator} (linear algebra), pylops real-FFT (signal
processing), and jax \texttt{linear\_transpose} (automatic
differentiation), disjoint from the 12 PUTs and the pre-registered
60-cell. Each library carries a real fixed defect in its adjoint code
path as a ground-truth anchor (scipy \#8900, pylops \#292, jax \#6223).
For each, the MR families are adj.self (self-adjointness) and adj.dual
(scaled/dual-operator transpose), and the mutant pool mixes
adjoint-breaking mutants (the real defect among them) with semantically
equivalent mutants a strong MR must \emph{not} kill; the real defects are
reproduced in extracted-diff form because the pre-fix releases are
uninstallable on the host (no \texttt{arm64} wheel). The arm is
confirmatory and additive: it does not alter the 60-cell denominator, the
primary-MP convention, or the pre-registered hypotheses. Full subject
specifications and fault mechanisms are in supplementary Appendix H;
results in Section~\ref{rq4-adjoint-extension-arm}.

\subsection{Empirical Study Procedure}\label{procedure-overview}

With the subjects, operators, and the adjoint extension arm fixed
(Sections~\ref{put-selection-12-puts-4-classes}--\ref{adjoint-extension-arm-subjects}), we now specify the procedure that produces each cell's
measurements.

\noindent\textsf{\small Pipeline order: Section~\ref{put-selection-12-puts-4-classes} PUTs $\to$ Section~\ref{cross-source-v4-protocol-summary} Mutant generation $\to$ LRCA L0 prescreen $\to$ Section~\ref{lrca-three-layer-diagnosis} Equivalence ($E_1 \wedge E_2$) $\to$ AVP $\to$ killed/survive $\to$ LRCA three-layer diagnosis $\to$ SMS + $C_1$\_share report.}

Each (i, k, j) cell is reported with \texttt{inst\_count},
\texttt{equiv\_count}, \texttt{killed\_count}, \texttt{survive\_count},
SMS\_\{i,k,j\}, C1\_share\_\{i,k,j\}, suspect\_share\_\{i,k,j\}, and the
three rates \texttt{inst\_rate\ /\ equiv\_rate\ /\ survive\_rate}
corresponding to RQ3.

\subsection{Cross-Source Generation Protocol}\label{cross-source-v4-protocol-summary}

To isolate the contributions of LLM same-source bias and MR-MP alignment
design to Cliff's delta, we ran a three-stage ablation:

{\def\LTcaptype{table} 
\begin{longtable}[]{@{}
  >{\raggedright\arraybackslash}p{(\linewidth - 6\tabcolsep) * \real{0.2500}}
  >{\raggedright\arraybackslash}p{(\linewidth - 6\tabcolsep) * \real{0.2500}}
  >{\raggedright\arraybackslash}p{(\linewidth - 6\tabcolsep) * \real{0.2500}}
  >{\raggedright\arraybackslash}p{(\linewidth - 6\tabcolsep) * \real{0.2500}}@{}}
\caption{Cross-source protocol.}\label{tab:p2-06}\\
\toprule\noalign{}
\begin{minipage}[b]{\linewidth}\raggedright
Version
\end{minipage} & \begin{minipage}[b]{\linewidth}\raggedright
Mutant pool
\end{minipage} & \begin{minipage}[b]{\linewidth}\raggedright
c-class primary MP
\end{minipage} & \begin{minipage}[b]{\linewidth}\raggedright
Use
\end{minipage} \\
\midrule\noalign{}
\endhead
\bottomrule\noalign{}
\endlastfoot
\textbf{v3} & Same-source (Claude Opus 4.6) & MP5 (pre-registered) & H2
baseline (pre-registered primary) \\
\textbf{v4} & \textbf{Cross-source} (Claude Opus 4.6 + GPT-5.4 +
DeepSeek chat) & MP5 (held at pre-registered) & Isolate LLM-source
diversity \\
\end{longtable}
}

The v4 protocol (\texttt{scripts/cross\_source\_campaign.py}) runs each
(PUT, operator) pair on Claude / GPT / DeepSeek with K=3 trials under an
identical prompt template (temperature 0.7, V1-V4 mechanical-validation
gate). Cross-source pool capacity: 37 operators $\times$ 3 sources $\times$ 3 trials =
333 attempts, 89.5\% V1-V4 pass, 298 confirmed mutants distributed nearly
equally (Claude 101, GPT 98, DeepSeek 99); final-stage rejection of
duplicate and QA-failing mutants leaves the 292 that enter the v4 pool
used for SMS and the AST audit.

\textbf{Declared confound: protocol asymmetry.} v3 used the
original Phase-1 dual-blind reviewer protocol (Claude generation +
GPT-5.4 review + DeepSeek arbitration); v4 passes V1-V4 mechanical gates
only and does \textbf{not} invoke a reviewer LLM. A fraction of the v3 $\to$
v4 source-axis contrast may therefore reflect a slight quality shift
rather than LLM-source diversity per se. The follow-up study will rerun
dual-blind on the v4 grid; full per-LLM token / latency / cost figures,
V1-V4 specifications, and the full differential-prompt protocol are in
\textbf{Appendix C.1}.

\subsection{Mutant-pool prescreen and equivalence detection}\label{mutant-pool-prescreen-and-equivalence-detection}

\textbf{Pool prescreen (LRCA L0).} Each candidate mutant passes three
gates:
\begin{enumerate}
\tightlist
\item Static lint and type checking.
\item A unit self-test on simple inputs to confirm that the PUT loads,
  runs, and returns a finite output.
\item A double-blind review sign-off under the Phase-1 protocol.
\end{enumerate} In Phase-1, the generator LLM is Claude Opus and the reviewer
LLM is GPT-5.4; the two roles are isolated, and the reviewer sees only
the PUT and the mutant code and outputs a (syntactic, executable,
fault-injected) triple. Inconsistent reviewer outputs go to a manual
arbitration queue (no more than 10\% of cases); double-confirmed mutants
enter the pool. We then randomly sample 20\% of the pool for manual
review by scientific-computing researchers, and downgrade the entire
batch to manual review if manual--reviewer inconsistency exceeds 10\%.
Each cell produces 30--50 candidates and retains 10--15 mutants. The v4
cross-source pool does \textbf{not} invoke the reviewer LLM, for cost
and speed reasons; a fraction of the v3 $\to$ v4 source-axis shift may
therefore reflect a quality decline rather than a source-diversity
contribution. We declare this protocol asymmetry in Section~\ref{cross-source-contributes-mutant-quality-not-effect-size};
full justification in Appendices C.2 and C.3).

\textbf{Equivalence detection (E1 $\wedge$ E2).} For each mutant
$s' \in \mathrm{mut}_j(S_i)$:

\begin{enumerate}
\tightlist
\item
  \textbf{E2 step.} Sample $K_{\mathrm{eq}} = 1000$ inputs
  $x \sim D_S$; compute
  $\|S_i(x) - s'(x)\|$; if $\leq \varepsilon_{\mathrm{eq}}$ for all
  samples, mark E2-pass.
\item
  \textbf{E1 step.} For every $mr \in \mathrm{MR}_{i,k}$, compare
  $\mathrm{AVP}(S_i, mr)$ with $\mathrm{AVP}(s', mr)$;
  if all consistent, mark E1-pass.
\item
  \textbf{Conjunction.} E1 $\wedge$ E2 $\to$ enter $\mathrm{equiv}_{i,k,j}$,
  exclude from SMS denominator. Otherwise pass to AVP-killed
  determination.
\end{enumerate}

E2 is run before E1 because E2 is computationally cheaper (K\_eq scalar
evaluations) and quickly discards numerically distinct mutants; the
remaining candidates are screened by AVP for MR-relation equivalence.
The joint condition is conservative: false-non-equivalent (E1 $\vee$ E2 fail)
is much easier than false-equivalent (both must mis-pass
simultaneously), biasing SMS slightly \emph{high}, an explicit
conservative engineering choice (Appendix A.3).

\textbf{Killed determination.} With OR-aggregation across mr in
MR\_\{i,k\}:

\[
\begin{aligned}
\mathrm{killed}(s', \mathrm{MR}_{i,k}) \iff{} & \exists\,mr \in \mathrm{MR}_{i,k}: \\
& \mathrm{AVP}(S_i, mr) = \text{pass} \wedge
  \mathrm{AVP}(s', mr) = \text{fail}.
\end{aligned}
\]

Each (i, k, j) cell is sampled N = 20 times (statistical replicate) to
compute the per-cell SMS, with mutant-level fail ratios feeding the LRCA
L1 decision. AVP version is pinned to the upstream infrastructure commit hash
(\texttt{\textless{}AVP-vX.Y\textgreater{}}); our reproducibility
package embeds the complete AVP source so that the package remains
self-consistent under any upstream evolution.

\subsection{Three-Layer Root-Cause Diagnosis}\label{lrca-three-layer-diagnosis}

For each killed mutant the three-layer decision tree applies in order:

\begin{itemize}
\tightlist
\item
  \textbf{L1 tolerance robustness} (all classes). N=20 replicates with a
  fail-ratio cutoff at 0.80; a kill that survives less than 80\% of the
  20 replicates is labelled C2 (numerical-tolerance perturbation) and
  exits.
\item
  \textbf{L2 OOD triage} (C / D classes only). Sample from
  $D_S^{\mathrm{valid}}$; if the mutant fails \emph{only} in the OOD
  region, label C3 (out-of-distribution) and exit.
\item
  \textbf{L3 statistical-assumption baseline} (B / D + Wilcoxon / DTW
  only). Pre-check IID / stationarity on the PUT's own repeated samples;
  if the AVP statistical assumption is itself violated, label C4 and
  exit.
\item
  \textbf{Artefact recheck}. An external reviewer re-examines mutant
  code + prompt history; if LLM / artefact evidence (e.g., mutator
  over-injection) is found, label C5; otherwise label C1.
\end{itemize}

Multi-label priority C5 \textgreater{} C4 \textgreater{} C3
\textgreater{} C2 \textgreater{} C1 takes the earliest confirmed
non-semantic cause (decision-tree control flow). Threshold calibration
over a 9-grid ($\texttt{ood\_band} \in \{0.02, 0.05, 0.10\}$ $\times$
$\texttt{tolerance\_multiplier} \in \{3.0, 10.0, 30.0\}$, repeats fixed
at 20) lifts H4 from 10/60 to 12/60 cells; the best combination
(\texttt{ood\_band\ =\ 0.02,\ tolerance\_multiplier\ =\ 3.0}) is
reported as primary, with default-threshold results retained as control
(\texttt{lrca\_60cell\_v3.json}). The calibration ceiling (12/60 = 20\%)
remains far below the 80\% pre-registered threshold; H4 is unattainable
on this dataset as an intrinsic property of LLM-mutant pools, not a
calibration issue. Detailed grid table, L0-L3 sub-protocols, and
decision-tree pseudocode are in \textbf{Appendix A.2 + C.4}.


\section{Results and Discussion}\label{statistical-analysis-and-empirical-results}

The empirical thresholds H1--H4 are stress tests of this particular
semantic-mutant instantiation, not prerequisites for the definition of
SMS. The positive validation target is narrower and more structural:
SMS must preserve the mutation-score denominator, separate
domain-semantic mutants from default first-order syntactic mutants, make
MR adequacy failures diagnosable, and distinguish mutant kill-rate,
semantic-stratum alignment, and real-defect detection. Thus the failed
thresholds below are reported as boundary findings about operator
applicability, MR design, and kill attribution; they do not invalidate
the metric construction.

\begin{figure}
\centering
\includegraphics[width=0.9\linewidth,height=\textheight,keepaspectratio,alt={60-cell heatmap of mean SMS over 12 PUTs \textbackslash times 5 MPs (cross-source primary). Diagonal cells (operator-MP aligned) show concentrated mass; off-diagonal mass reflects bleed into adjacent MPs. Empty rows = PUT-level zero-mass cohort (PUTs A1, B1, D2; see Section~\ref{decoupling-between-r_sem-and-r_kill}).}]{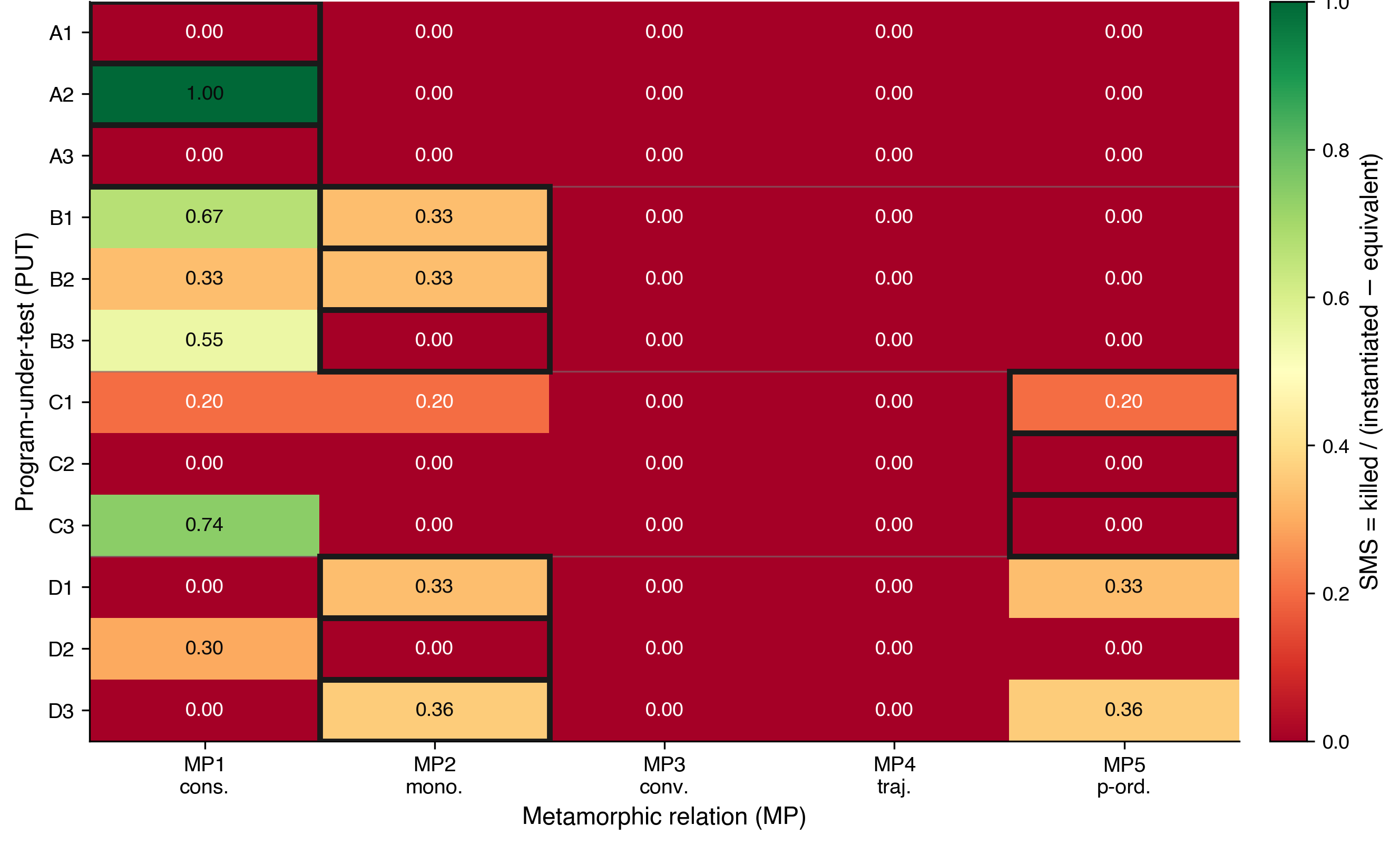}
\Description{60-cell heatmap of mean SMS over 12 PUTs by five meta-patterns under the cross-source primary pool. Aligned cells concentrate the main signal, while many cross cells are zero; PUTs A1, B1, and D2 form all-zero rows.}
\caption{60-cell heatmap of mean SMS over 12 PUTs \(\times\) 5 MPs (v4
cross-source primary). Diagonal cells (operator-MP aligned) show
concentrated mass; off-diagonal mass reflects bleed into adjacent MPs.
Empty rows = PUT-level zero-mass cohort (PUTs A1, B1, D2; see
Section~\ref{decoupling-between-r_sem-and-r_kill}).}\label{fig:heatmap}
\end{figure}

\subsection{Statistical pipeline}\label{statistical-pipeline}

The primary statistical reporting follows a pre-registered hierarchy:

{\def\LTcaptype{table} 
\begin{longtable}[]{@{}
  >{\raggedright\arraybackslash}p{(\linewidth - 4\tabcolsep) * \real{0.3333}}
  >{\raggedright\arraybackslash}p{(\linewidth - 4\tabcolsep) * \real{0.3333}}
  >{\raggedright\arraybackslash}p{(\linewidth - 4\tabcolsep) * \real{0.3333}}@{}}
\caption{Statistical pipeline.}\label{tab:p2-07}\\
\toprule\noalign{}
\begin{minipage}[b]{\linewidth}\raggedright
RQ
\end{minipage} & \begin{minipage}[b]{\linewidth}\raggedright
Primary statistics
\end{minipage} & \begin{minipage}[b]{\linewidth}\raggedright
Reporting format
\end{minipage} \\
\midrule\noalign{}
\endfirsthead
\endhead
\bottomrule\noalign{}
\endlastfoot
RQ3 & inst\_rate, equiv\_rate, C1\_share, survive\_rate & 60-cell
heatmap + 4-class marginals \\
RQ4 & aligned-SMS vs cross-SMS; sparse $\circ$ vs dense \textbullet\textbullet equiv\_rate &
Cliff's delta + odds ratio + 95\% bootstrap CI \\
RQ5 & $\Delta$SMS\_c (c $\in$ \{A,B,C,D\}); CV($\Delta$SMS) & sign test (df = 3) +
descriptive forest plot \\
RQ5 (desc.) & Spearman $\rho$ + Kendall $\tau$ (SMS vs PC) & scatter + dual-metric ranking
comparison \\
\end{longtable}
}

This pipeline answers the empirical questions RQ3--RQ5; RQ1 is answered
by Theorem 9.1 (Section~\ref{sms-ms-degeneration-formal-statement}) and RQ2 by the AST-normalised structural audit
(Section~\ref{p2-vs-syntactic-mutants-12-put-empirical-layer-3}), outside the statistical pipeline.

Where multiple per-class tests are reported (the per-class Friedman tests
of Section~\ref{rq3-cross-class-consistency-and-friedman}), they carry a Bonferroni correction across the four classes; we
do not report a separate cell-level family of p-values, so no family-wise
cell-level correction is claimed. Bootstrap 95\% CIs use 1,000 iterations
as default (B = 10,000 for the headline H2 delta CI). Mixed-effects
modelling
(\texttt{sms\ \textasciitilde{}\ C(class)\ *\ C(operator)\ +\ (1\ \textbar{}\ put)})
failed with a Singular matrix at N = 60: the class $\times$ operator
interaction design matrix is rank-deficient and the PUT random-intercept
variance collapses to the boundary at this sample size. The Friedman test
is the non-parametric formal alternative. The four hypotheses are
pre-registered: H1 (operator implementability) requires $\geq$ 4 of 5
operators producing $\geq$ 5 non-equivalent mutants on $\geq$ 9/12 PUTs; H2
(aligned-vs-cross) requires odds ratio $\geq$ 3.0 \emph{and} Cliff's delta $\geq$
0.474 (Romano 2006 large-effect threshold); H3 (cross-class consistency)
requires sign-test 4/4 across 4 classes plus CV($\Delta$SMS) \textless{} 0.5;
H4 (LRCA mass) requires mean suspect\_share $\leq$ 0.20. SMS variants
(SMS, SMS\_unfiltered) and construction lemmas are in
\textbf{Appendix D.1}.

\subsection{RQ3: Semantic-Mutation Score Distribution}\label{rq1-60-cell-distribution}

{\def\LTcaptype{table} 
\begin{longtable}[]{@{}ll@{}}
\caption{RQ3: Semantic-mutation score distribution.}\\
\toprule\noalign{}
Metric & Value \\
\midrule\noalign{}
\endhead
\bottomrule\noalign{}
\endlastfoot
Number of cells & 60 \\
Mean SMS & 0.104 \\
Median SMS & 0.000 \\
Std SMS & 0.213 \\
Cells with SMS = 0 & 45 / 60 \\
Mean mutants / cell (v4) & 24.3 (range 10-30) \\
\end{longtable}
}

The \textbf{zero-mass dominance} (45/60 = 75\%) is concentrated in the
cross-MP slice: \textasciitilde88\% (42/48) of cross cells are zero,
\textasciitilde25\% (3/12) of aligned cells are zero. Cliff's delta is
well-defined under this distribution but its inference is effectively
dominated by
\texttt{n\_aligned\ =\ 12\ +\ n\_cross\_nonzero\ $\approx$\ 6\ =\ 18}, not the
surface n=60. Median odds ratio is formally infinite (median(cross) =
0); we report ``aligned median \textgreater{} 0 = cross median'' as
auxiliary qualitative evidence.

\textbf{Effective-n note.} The surface n\_aligned = 12 and n\_cross = 48
mask a much smaller information content: only about 18 observations carry
nonzero SMS (12 aligned + $\approx$ 6 nonzero cross). This is a heuristic
information count, not a formal effective sample size, but it explains the
wide 95\% bootstrap CI {[}0.127, 0.740{]} (upper/lower ratio $\approx$
5.83, consistent with the known liberal tendency of the percentile
bootstrap when few observations carry signal). The implications for power and the H2 verdict direction are
quantified jointly with the stipulated-alternative analysis in Section~\ref{rq2-stipulated-alternative-power}.
PUT-class diversification at n $\geq$ 30 is a testable route to relaxing
the effective-n constraint. Power and effect-size disambiguation are
treated jointly with the stipulated-alternative analysis in Section~\ref{rq2-stipulated-alternative-power}
(avoiding repetition).

\textbf{Consistency with LLM-mutant literature.} \citet{tip2025llmorpheus}
LLMorpheus reports high cross-MP failure proportions on JavaScript
scientific-computing PUTs (specific numbers not listed in cited
literature); zero-mass dominance appears to be a shared characteristic
of LLM-mutant + existing MR-MP alignment designs, not a quirk of this
paper's PUT selection.

\textbf{H1 verdict: not met.} H1 (RQ3) requires at least 4 of the 5
operators to produce $\geq 5$ non-equivalent (confirmed) mutants on at
least 9 of the 12 PUTs. Table~\ref{tab:p2-32} counts confirmed
non-equivalent mutants per operator class across the 12 PUTs: only the
Hyperparameter operator clears the $\geq 9/12$ bar.

{\def\LTcaptype{table} 
\begin{longtable}[]{@{}ll@{}}
\caption{H1: PUTs reaching $\geq 5$ non-equivalent mutants, per operator
class.}\label{tab:p2-32}\\
\toprule\noalign{}
Operator class & PUTs with $\geq 5$ non-equiv mutants (of 12) \\
\midrule\noalign{}
\endfirsthead
\endhead
\bottomrule\noalign{}
\endlastfoot
Conservation Erosion (CE) & 4 / 12 \\
Operator Substitution (OS) & 5 / 12 \\
Hyperparameter (HP) & \textbf{9 / 12} \\
Trajectory Flip (TF) & 5 / 12 \\
Structural Injection (SI) & 1 / 12 \\
\end{longtable}
}

Because the semantic operators are class-targeted, each is applicable
to a subset of the four PUT classes rather than uniformly to all twelve.
Only the broadly-applicable Hyperparameter operator clears the 9-PUT
threshold. Thus 1 of 5 operators meets the criterion and \textbf{H1 is
not met as pre-registered}; the uniform-applicability assumption behind
the $\geq 9/12$ bar does not hold for class-targeted operators.

As a secondary interpretive denominator, we also count only PUTs for
which the operator family has an explicit class-specific specialization
in Appendix B.2 and the operator-level pilot records an attempted
instantiation in Appendix C.4. Under this applicability-adjusted view,
CE reaches 4/8 applicable PUTs, OS 5/7, HP 9/9, TF 5/6, and SI 1/6. The
secondary denominator does not overturn the pre-registered H1 verdict,
but it identifies the scientific source of the failure: HP and TF are
implementable across their intended classes, OS is moderately broad, and
SI remains a narrow high-risk family rather than a uniformly applicable
operator. The per-operator implementability profile ($r_{\mathrm{sem}}$,
$d_{\mathrm{impl}}$) is reported in Appendix C.4.

\subsection{RQ4: Aligned and Cross-Pattern Comparisons}\label{rq2-aligned-vs-cross-cliffs-delta}

{\def\LTcaptype{table} 
\begin{longtable}[]{@{}llll@{}}
\caption{RQ4: Aligned and cross-pattern comparisons.}\label{tab:p2-09}\\
\toprule\noalign{}
Slice & n & Mean SMS & Median SMS \\
\midrule\noalign{}
\endhead
\bottomrule\noalign{}
\endlastfoot
aligned (j = k) & 12 & 0.275 & 0.267 \\
cross (j $\neq$ k) & 48 & 0.061 & 0.000 \\
\end{longtable}
}

Two delta point-estimates under the pre-registered primary-MP
convention:

\begin{itemize}
\tightlist
\item
  \textbf{v3 (same-source, pre-registered):} delta = \textbf{0.323},
  95\% CI {[}0.017, 0.622{]}
\item
  \textbf{v4 (cross-source, c-class held at MP5):} delta =
  \textbf{0.314}, 95\% CI {[}0.014, 0.622{]}. The cross-source pool
  aggregates Claude / GPT / DeepSeek under an identical prompt; the
  c-class primary is held at the pre-registered MP5 so that the v3 $\to$ v4
  contrast holds the prompt and primary-MP choice fixed. Because v4 uses
  mechanical gates rather than the v3 dual-blind reviewer step, the
  source-axis reading is qualified by the protocol-asymmetry analysis in
  Section~\ref{r13-protocol-asymmetry-magnitude-estimate}.
\end{itemize}

\textbf{H2 verdict: not met.} H2 carries two pre-registered criteria,
Cliff's $\delta \geq 0.474$ and an aligned-to-cross odds ratio
$\geq 3.0$. Neither delta crosses the \citet{romano2006cliff}
large-effect threshold 0.474, 0.323 and 0.314 both fall just below the
\citet{romano2006cliff} medium threshold of 0.330, so the delta criterion fails. The
odds ratio is degenerate: because the cross-slice median SMS is exactly
0 (table above), the median odds ratio is $+\infty$, so the ratio
criterion is satisfied only trivially under zero-inflation and carries no
evidential weight. H2 is therefore not met as a large-effect target. The
contrast $\delta_{\mathrm{v4}} - \delta_{\mathrm{v3}} = -0.009$ (95\% CI covers
zero) is too small to interpret as a stable LLM-source-diversity effect
under the present asymmetric protocol: replacing a same-source pool with
a three-LLM cross-source pool does not appreciably shift Cliff's delta
within this design, but the exact source-diversity contribution requires
a dual-blind v4 rerun. We frame H2 as an underpowered exploratory
contribution and defer full verification to a follow-up study with a
larger PUT sample.

This medium-effect placement is consistent with \citet{tip2025llmorpheus}
LLMorpheus's medium-effect range on JavaScript LLM mutants, a contextual
literature observation (estimand caveat: their delta compares ``LLM vs
traditional mutants on fault detection'', ours compares ``aligned vs
cross MP slice within one pool''; the numerical similarity is not
substantive support).

\subsection{RQ4: Stipulated-Alternative Power Analysis}\label{rq2-stipulated-alternative-power}

The Section~\ref{rq1-60-cell-distribution} effective-n constraint motivates an explicit power analysis. A
plug-in bootstrap (5,000 replications, seed = 42) samples with
replacement from the observed (n = 12, n = 48) v4 SMS distributions. The
bootstrap exceedance-probability table (probability that the resampled
$\hat\delta$ exceeds each threshold under the observed distribution, not
power against a stipulated alternative) is:

{\def\LTcaptype{table} 
\begin{longtable}[]{@{}lll@{}}
\caption{RQ4: Bootstrap exceedance probability under the observed
distribution.}\\
\toprule\noalign{}
Threshold & Interpretation & Exceedance prob.\ at (12, 48) \\
\midrule\noalign{}
\endhead
\bottomrule\noalign{}
\endlastfoot
$\delta$ \textgreater{} 0.000 & Any effect & \textbf{0.997} \\
$\delta$ \textgreater{} 0.147 & Small & 0.966 \\
$\delta$ \textgreater{} 0.330 & Medium & 0.759 \\
\textbf{$\delta$ \textgreater{} 0.474} & \textbf{Large (H2)} &
\textbf{0.423} \\
\end{longtable}
}

The plug-in result answers the question ``given the \emph{observed}
distribution, how often do we exceed the threshold?'' A
stipulated-alternative simulation answers a more pointed question: ``if
the \emph{truth} equals the H2 boundary 0.474, how often does the (12,
48) design return $\delta \geq 0.474$?'' SMS distributions have heavy ties at zero
(45 of 60 cells), so a raw shift is discontinuous: any $\varepsilon$ \textgreater{}
0 jumps $\delta$ from 0.314 to 0.74. We therefore use a mixture, drawing the
aligned sample with probability w from (observed\_aligned + 0.001) and
with probability 1 − w from observed\_aligned, calibrating w = 0.094 to
realise $E[\delta] = 0.4746$.

{\def\LTcaptype{table} 
\begin{longtable}[]{@{}lll@{}}
\caption{RQ4: Stipulated-alternative power.}\label{tab:p2-11}\\
\toprule\noalign{}
Stipulated truth & Criterion & Power \\
\midrule\noalign{}
\endfirsthead
\endhead
\bottomrule\noalign{}
\endlastfoot
0.474 & $\hat{\delta} \geq 0.474$ (point estimate) & \textbf{0.491} \\
0.474 & Lower 95\% CI bound \textgreater{} 0 (any effect) & 0.868 \\
\end{longtable}
}

Even when the truth equals the H2 boundary, this design returns ``not
met'' verdicts in roughly half of replications. This supports the
framing in Section~\ref{rq2-aligned-vs-cross-cliffs-delta}: the H2 verdict is a factual statement about the point
estimate failing to clear the threshold, not a claim that the effect is
necessarily smaller than 0.474. Increasing the sample size narrows the
confidence interval but cannot lift the point estimate. The plug-in
sample-size sweep (n\_aligned $\in$ \{6, 12, \ldots, 60\}; n\_cross = 4 $\times$
n\_aligned; power for $\delta$ \textgreater{} 0 reaches 0.974 at n\_aligned = 6
and 0.996 at 12, then plateaus) is in Appendix D.3.

\textbf{The source-axis contrast is underpowered too.} We did not run a
separate power simulation for the source-axis contrast $\Delta\delta$ =
−0.009 (CI covers zero); its estimand and dependence structure differ
from the aligned-vs-cross $\delta$, so the 49.1\% figure above does not
transfer to it. The small design (n = 12 PUTs) leaves the −0.009 null
consistent with a wide range of true source-diversity effects, so we
explicitly do not read it as evidence that source diversity is inert. A
properly-powered strong-sense test is deferred to a follow-up study.

\textbf{Note on monotone-transformation invariance.} Cliff's $\delta$ is
rank-based: a function of $U / (n_1 \cdot n_2)$, so applying a logit transform to
SMS gives $\delta_{\mathrm{logit}} \equiv \delta_{\mathrm{raw}}$ by construction. This is consistent with the
rank-invariance theorem and does \textbf{not} constitute additional
robustness evidence. The genuine robustness threats to the H2 verdict
come from the zero-mass dominance (Section~\ref{rq1-60-cell-distribution}), not from any metric-scale
choice.

\subsection{RQ5: Cross-Class Consistency}\label{rq3-cross-class-consistency-and-friedman}

{\def\LTcaptype{table} 
\begin{longtable}[]{@{}lll@{}}
\caption{RQ5: Cross-class consistency and Friedman statistics.}\label{tab:p2-12}\\
\toprule\noalign{}
Class & Mean SMS (v3) & Mean SMS (v4) \\
\midrule\noalign{}
\endfirsthead
\endhead
\bottomrule\noalign{}
\endlastfoot
a (numeric) & 0.067 & 0.067 \\
b (probabilistic) & 0.156 & 0.148 \\
c (surrogate) & 0.047 & \textbf{0.089 (+91.5\%)} \\
d (ML) & 0.081 & 0.112 (+38.3\%) \\
\end{longtable}
}

Sign test (within-class aligned mean − cross mean, sign = +): \textbf{3
/ 4 (partial) under both v3 (same-source) and v4 (cross-source)}; the
b-class is the inverted cell.

\textbf{H3 verdict: not met.} H3 carries two pre-registered criteria, a
within-class sign test of 4/4 and CV($\Delta$SMS) $<$ 0.5. The sign test
is 3/4 (partial) under both v3 and v4, cross-source pooling does not
flip the b-class inversion, so the consistency criterion fails. The
per-class effects $\Delta$SMS\_c (aligned mean $-$ cross mean) span from
negative in the inverted class to large positive elsewhere, giving
CV($\Delta$SMS) well above 1, far exceeding the 0.5 dispersion threshold,
so the dispersion criterion also fails.

The mixed-effects primary model
\texttt{sms\ \textasciitilde{}\ C(class)\ *\ C(operator)\ +\ (1\ \textbar{}\ put)}
returned Singular matrix; the fallback
\texttt{sms\ \textasciitilde{}\ C(class)\ +\ C(operator)\ +\ (1\ \textbar{}\ put)}
had PUT random-intercept variance hit boundary 0 (degenerate). We
therefore use Friedman as the non-parametric formal alternative:

\begin{itemize}
\tightlist
\item
  \textbf{PUT (n=12) $\times$ MP (k=5): chi\^{}2 = 15.30, p = 0.0041}
  (significant).
\item
  MP rank means: 2.92, 2.58, 2.08, 3.08, 4.33.
\item
  Per-class Friedman with \textbf{Bonferroni $\times$ 4 correction}: a
  1.000 / b \textbf{0.116} / c 1.000 / d 1.000, no per-class
  result remains significant after correction. Kendall's W effect sizes:
  a 0.333 (small) / b 0.898 (large concordance, but caveat: N=3 per
  class makes this label nominal only) / c 0.333 / d 0.417.
\end{itemize}

\textbf{Caveat.} Friedman tests ``are MP rank differences present''
(averaged over PUTs); H3 tests ``is direction consistent across 4
classes''. These are logically independent. H3 stands on the Section~\ref{rq3-cross-class-consistency-and-friedman} sign
test; per-class Friedman is descriptive only (small N=3 per class).

\subsection{RQ5: Semantic-Mutation Score and Pattern Coverage}\label{rq4-sms-vs-pattern-coverage}

\textbf{Status.} The SMS--Pattern-Coverage relationship, a descriptive
sub-question of RQ5, is reported as a descriptive observation, not a
hypothesis test, because n = 12 places the 95\% Spearman CI at roughly
{[}−0.5, +0.6{]}, so the test cannot distinguish zero,
moderate-positive, or moderate-negative correlation. The numbers below
are recorded so that a future study (n $\geq$ 30 PUTs) can pre-register a directional
hypothesis.

Pattern Coverage (PC) per PUT = \#triggered (MP\_k, R\_outcome) cells /
10. Range {[}0.500, 1.000{]}, mean 0.733. Pairing with mean SMS over 5
MPs: Spearman rho = \textbf{0.163} (p = 0.613); Kendall tau = 0.136 (p =
0.568); n = 12.

\textbf{Statistical-power caveat first.} At n = 12, Spearman's 95\% CI
is approximately {[}-0.5, +0.6{]} (Fisher-z approximation, a rough
small-sample heuristic); the test cannot distinguish ``zero
correlation'' from ``moderate positive correlation'' or ``moderate
negative correlation''. p = 0.61 / 0.57 does \textbf{not} support the
strong claim that ``SMS is independent of PC'' or the strong claim that
``SMS is strongly correlated with PC''. The conservative finding is
\emph{no detectable correlation at n = 12}; orthogonality is a
hypothesis for a future study (n $\geq$ 30 PUTs or refined PC operationalisation; full PC
operationalisation in Appendix D.5).

\subsection{RQ4: Root-Cause Diagnostic Results}\label{h5-lrca-mass}

\begin{table}[t]
\caption{RQ4: Root-cause diagnostic mass.}
\label{tab:p2-13}
\centering
\begin{tabular}{ll}
\toprule
Metric & Value \\
\midrule
Mean C1\_share (default threshold) & 0.164 \\
Mean C1\_share (calibrated best, ood\_band = 0.02) & \textbf{0.209} \\
Mean suspect\_share (calibrated best) & \textbf{0.791} \\
Cells with suspect\_share $\leq$ 0.20 (secondary) & 12 / 60 = 20.0\% \\
\bottomrule
\end{tabular}
\end{table}

\textbf{H4 verdict: not met.} The pre-registered criterion is mean
suspect\_share $\leq$ 0.20; the observed mean is \textbf{0.791} (v4),
decisively above the threshold, so H4 fails on its own metric. A dense
cutoff sweep (Appendix D.2) confirms this is an intrinsic property of
LLM-mutant pools rather than a calibration artefact: the secondary
per-cell pass-ratio (cells with suspect\_share $\leq$ 0.20) is flat at
12 / 60 = 20\% across cutoffs $\in$ {[}0.05, 0.40{]}, because v4's
suspect\_share distribution is severely bimodal (median 1.0, with 48
cross cells near 1 and 12 aligned cells near 0; the {[}0.20, 0.80{]}
interior is nearly empty). H4 was pre-registered before pool
characteristics were known, so this is a finding worth reporting.

\textbf{C1-weighted sensitivity.} Because LRCA deliberately does not
filter SMS, we add a diagnostic secondary view
$\mathrm{SMS}_{\mathrm{C1}} = \mathrm{SMS} \times \mathrm{C1\_share}$.
On the v4 60-cell table, mean SMS is 0.104 and mean
$\mathrm{SMS}_{\mathrm{C1}}$ is 0.0873; nonzero cells fall from 15/60 to
13/60. The conclusion is therefore conservative: suspect mass is high
and H4 is not met, but the main positive SMS signal is not created solely
by suspect labels. We keep raw SMS as the primary metric because LRCA is
a diagnostic attribution layer, not part of the mutation-score
denominator.

\subsection{RQ4: Strong-Boundary Failure Modes}\label{rq4-strong-boundary}

The 60-cell audit (Sections~\ref{rq1-60-cell-distribution}--\ref{h5-lrca-mass}) exercises strong MRs in the regime where
Theorem 2 holds. To map the strong boundary of Proposition 2 we run a
complementary study on six third-party or textbook programs spanning the
meta-patterns, kept separate from the 12 PUTs. Wherever a genuine fixed
defect exists in a third-party library we use real version-switching:
the pre-fix and post-fix releases run unmodified in isolated environments,
and we record the repository, the version pair, and the issue/PR, and
we fall back to a textbook faithful implementation only when the real
defect cannot be exercised on the host (no \texttt{arm64} wheel, a
heavyweight Fortran build, or a drift below machine resolution).
$\varepsilon_{\mathrm{tol}}$ is an explicit factor in every case.

{\def\LTcaptype{table} 
\begin{longtable}[]{@{}p{1.6cm}p{3.4cm}p{2.5cm}p{2.6cm}p{1.7cm}@{}}
\caption{RQ4: Strong-boundary map over six programs.}\label{tab:p2-30}\\
\toprule\noalign{}
Family & Program (provenance; method) & Strong MR ($\varepsilon_{\mathrm{tol}}$) & Correct / faulty verdict & Boundary role \\
\midrule\noalign{}
\endfirsthead
\endhead
\bottomrule\noalign{}
\endlastfoot
inv.con & non-conservative Burgers; \citep{hou1994nonconservative} (textbook) & shock speed ($0.03$) & $0.498$ pass / $-0.002$ killed & strong \\
inv.eqv & scipy \texttt{align\_vectors} $1.13.0\to1.13.1$, \#20660 (real) & rotation residual ($10^{-6}$) & $10^{-16}$ pass / $2.0$ killed & strong \\
adj.self & scipy \texttt{\_adjoint} \#8900/PR\#8962 (real, extracted-diff) & adjoint-identity residual & $10^{-16}$ pass / $1.4$--$2.0$ killed & strong \\
rev.traj & leapfrog vs explicit Euler; \citep{hairer2006geometric} (textbook) & energy-drift ratio ($\leq 3$) & $1.00$ pass / $1952$ killed & strong \\
PINN & soft vs hard BC; DeepXDE \#26 (real torch) & exact BC $|u(0)|$ ($10^{-4}$) & hard $0$ pass / soft $5\times10^{-4}$ killed & FP (weak MR) \\
struct.\ (FN) & numpy float32 RNG $1.21.6\to1.22.0$, \#17478 (real) & mean / range / dist & both pass; buggy survives & FN (surviving) \\
\end{longtable}
}

\textbf{Strong regime (four cases).} For inv.con, inv.eqv, adj.self, and
rev.traj the correct program satisfies the strong MR within
$\varepsilon_{\mathrm{tol}}$ while the faulty program is killed, exactly
as Theorem 2 predicts. Two are real fixed library defects (scipy
\texttt{align\_vectors}; scipy \texttt{\_ScaledLinearOperator.\_adjoint},
whose pre-fix release has no \texttt{arm64} wheel and is therefore
exercised by applying the exact fix diff in reverse); two are textbook
contrasts standing in for a real defect that cannot be exercised on the
host, a conservative versus non-conservative Burgers scheme for the
GeoClaw \texttt{filpatch} conservation bug, and a symplectic leapfrog
versus explicit Euler integrator for the REBOUND IAS15 energy drift, whose
real signature near $10^{-14}$ lies below resolution.

\textbf{False positive, weak MR (PINN).} A physics-informed network
that enforces the boundary condition as a soft penalty is a legitimately
trained, non-faulty model, yet its exact-boundary residual
$|u(0)| \approx 5\times10^{-4}$ exceeds a strict
$\varepsilon_{\mathrm{tol}} = 10^{-4}$, so the exact-boundary strong MR
flags it: the discretisation (the soft constraint), not a fault, breaks
the invariant. This is the discrete-program instance of the
rotational-symmetry counterexample, a continuous invariant reproduced
only approximately, and the empirical face of Proposition 2(i). The
hard-constraint ansatz $u = x(1-x)\,\mathrm{NN}(x)$ satisfies the invariant
by construction and passes, confirming that the violation is a property of
the discretisation, not of training.

\textbf{False negative, surviving non-equivalent mutant (RNG).} The
numpy float32 generator defect (\#17478) zeroes the lowest mantissa bit of
every sample; the structural MR family, mean $\approx 0.5$, range
$[0,1)$, distribution shape, is satisfied by both releases within
$\varepsilon_{\mathrm{tol}}$, so the genuinely non-equivalent buggy version
survives. This is coincidental satisfaction of the MR, and because the
fault signature lies below the structural tolerance it is a tolerance
fault; the surviving mutant is non-equivalent (an independent
odd-bit-fraction discriminator reads $0.0$ for the buggy release versus
$\approx 0.5$ for the correct one), and therefore distinct from a
classical equivalent mutant. It is the empirical face of Proposition
2(ii).

\textbf{The $\varepsilon_{\mathrm{tol}}$ trade-off.} The PINN case makes
the sensitivity--specificity coupling runnable: at the strict
$\varepsilon_{\mathrm{tol}} = 10^{-4}$ the legitimately-trained soft-BC
model is killed (a false positive), while at a loose
$\varepsilon_{\mathrm{tol}} = 10^{-3}$ it survives correctly; if the same
residual pattern instead arose from a true boundary-condition fault,
loosening $\varepsilon_{\mathrm{tol}}$ would turn that survival into a
false negative. Tightening
$\varepsilon_{\mathrm{tol}}$ trades false negatives for false positives and
loosening it does the reverse, as Proposition 2 requires. The strong
boundary is therefore the honest scope of the duality, a tunable
operating point, not a defect to be removed.

\subsection{RQ4: Adjoint Extension Evidence}\label{rq4-adjoint-extension-arm}

Where Section~\ref{rq4-strong-boundary} maps the boundary at which the duality fails, a small
confirmatory arm checks it holding in the forward direction for the sixth
meta-pattern $\psi_6$. On the three third-party linear-operator libraries
of Section~\ref{adjoint-extension-arm-subjects} (scipy, pylops, jax), each carrying a real fixed adjoint defect, a
strong adjoint MR detects only the active mutants: across all three
substrates the correct program survives, every non-equivalent mutant
(including the three real defects scipy \#8900, pylops \#292, and jax
\#6223) is killed, and no semantically equivalent mutant is killed, giving
a forward SMS of 1.00 with zero false positives on the equivalent set
(full subjects, mutant pools, and the per-substrate kill table in
supplementary Appendix H). The arm is confirmatory and additive, it
does not enter the 60-cell denominator and leaves H1--H4 unchanged, and
carries two scope caveats: the scipy \#8900 defect is shared with the Section~\ref{rq4-strong-boundary}
boundary study (so it is not counted as independent corroboration twice),
and the pools are hand-constructed, low-cardinality sets with the real
defects reproduced in extracted-diff form, so the arm demonstrates rather
than measures adequacy (adjoint-arm reproduction-fidelity threat).


\phantomsection\label{discussion}

\subsection{Cross-source contributes mutant quality, not effect size}\label{cross-source-contributes-mutant-quality-not-effect-size}

Going from same-source to cross-source under the pre-registered
primary-MP convention shifts Cliff's $\delta$ by only −0.009 (95\% CI covers
zero). Cross-source pooling raises mean C1\_share from 0.164 to 0.209 (a
27\% relative increase), class-c mean SMS by \textbf{+91.5\%}, and
class-d mean SMS by 38.3\%. Under an identical prompt template, three
LLMs produce similar distributions for the aligned-vs-cross question.
Within this fixed-prompt design, LLM source identity is therefore not the
dominant factor behind the H2 ceiling: cross-source pooling improves the
diagnostic quality of the kill set (LRCA C1 share, class-mean SMS)
without moving the aligned-vs-cross effect size. This is a scoped null,
not a general claim that LLM diversity is inert. The strong-sense
source-diversity test, with per-LLM differential prompts (V\_persona,
V\_cot), is deferred to a follow-up study. Appendix C.1 records the full
protocol.

The Section~\ref{p2-vs-syntactic-mutants-12-put-empirical-layer-3} evidence (5.14\% AST overlap, HP/SI/TF at 0/0/0) confirms that
the medium-effect ceiling is not an artefact of LLM-pool overlap with
the syntactic-mutant space. 94.86\% of v4 mutants are AST-disjoint from
cosmic-ray defaults, and the three classes unreachable under default
first-order configurations (HP, SI, TF; 159 of 292 mutants; HOM not
refuted, see Section~\ref{preventive-defence-framing}) lie outside that space by construction. A
larger aligned-versus-cross effect therefore points first to MR-design
refinement. A larger sample would mainly reduce uncertainty around this
design unless the MR design itself changes.

\textbf{A numerical-coincidence note.} \citeauthor{1929c365d171c78c9c24b242b5b57e2832bc907b}'s \citeyearpar{1929c365d171c78c9c24b242b5b57e2832bc907b}
\textasciitilde20\% productive-mutant rate at Google is close to our
calibrated C1\_share of 0.20. Their construct is a developer survey
(subjective usefulness), and LRCA's C1 is a diagnostic annotation over
observed kill behaviour; the agreement between the two is contextual,
not mechanism validation.

\subsection{Decoupling between Semantic Feasibility and Kill Rate}\label{decoupling-between-r_sem-and-r_kill}

The operator-level pilot in Appendix C.4.1 reveals a sharp pattern: HP,
TF, and OS operators on the c-class (surrogate) and d-class (machine
learning) PUTs reach R\_sem $\approx$ 1.0 (semantic feasibility, the LLM
successfully produces a syntactically valid, executable, intent-bearing
mutant) but R\_kill = 0 under the PUT's primary MP (the AVP fails to
detect the mutant under that MR). Concretely, six cells show R\_sem $\geq$
0.9 with R\_kill = 0: c1\_CE1 (GPR \texttt{noise\_level} 1e-4 $\to$ 1e-1),
c2\_OS1 (polynomial $\to$ spline basis), c3\_HP1 (relu $\to$ tanh), c3\_TF1
(max\_iter 1000 $\to$ 5), d1\_HP1 (MLP $\alpha$ 1e-4 $\to$ 1.0), and d3\_HP1 (LR C 1.0
$\to$ 1e-4). In each, the mutant \emph{is} a valid hyperparameter-semantic
injection, but the primary MP, MP5 asymptotic for the c-class and MP2
monotonicity for the d-class, is an asymptotic or statistical relation,
and the parameter deviation it tolerates is wide enough to absorb the
mutant.

The cell-level evidence in Section~\ref{rq1-60-cell-distribution} reproduces this pattern: 75\% of cells
are zero, concentrated in the cross slices. Under v4 cross-source, mean
C1\_share rises from 0.164 to 0.209 (+27\%): among the few killed
mutants, the cross-source pool produces cleaner LRCA diagnostic labels.
Source diversity therefore improves the \emph{quality} of the kill set
without expanding its \emph{coverage}. \textbf{The engineering insight is that
operator-MP alignment in MR design is necessary for strong SMS signals;
merely enlarging the semantically feasible mutant pool dilutes the C1
proportion without lifting the kill rate.} Two open questions remain:
whether SMS can infer which MP class is missing for which PUT class, and
whether LRCA can be calibrated with class-specific thresholds separating
likely semantic failures from artefacts.

This decoupling motivates the source-axis reading in Section~\ref{rq2-aligned-vs-cross-cliffs-delta}: the v3 $\to$
v4 micro-shift of −0.009 estimates LLM source diversity under a fixed
prompt, while MR-design adequacy at the operator-MP level governs the
kill rate in this design.

\textbf{Real-defect evidence slice.} A 34-case real-defect evidence
slice gives a useful check on this interpretation without changing the
paper into a benchmark article. We report verified result totals only:
defect mining, rejected cases, repository tooling, and benchmark release
are left outside the paper. The result-level evidence summary used here is included
in the supplementary evidence package: it records the five-stratum counts,
the aggregate mutation-phase statistics, the 30/30 real-defect face, and
the non-nesting case identifiers, but not candidate mining or tooling. In
this mutation-phase slice, the
pattern-derived MR group beats
literature-generic baselines under the primary all-mutant denominator
(mean paired difference +0.101, Holm-adjusted \(p=0.046\), Cliff's
\(\delta=+0.247\)), but the advantage is modest: the pattern-derived
group does not significantly dominate its ablations under the same
denominator. On the real-defect face, however, the distinction becomes
sharper: the pattern-derived relation detects every real defect in the
reported mutation-suite face (30/30), whereas generic and random
relations are mostly blind, and ablations that look close in aggregate
kill-rate often lose the real defect.

The slice was selected by a fixed inclusion rule: a case had to have a
verified semantic fault, an executable pre/post or faithful contrast, and
an MR-detectable behavioral difference under the declared stratum.
Cases requiring unsupported build stacks, lacking an executable contrast,
or failing MR-detectability verification were excluded before the
reported mutation-phase comparison. No case was dropped after observing
the comparative SMS or baseline detection results. The 30/30 figure is
therefore a face-level check on construct separation, not a corpus-level
claim about real-defect coverage.

This pattern is consistent with the construct separation predicted by the
fiber view of Section~\ref{theory-fibers}. A mutant kill-rate is evidence about which sampled
semantic effects an MR observes; it is not, by itself, a verdict on MR
quality. The same real-defect slice also records non-nesting cases in
which generic baselines kill mutants that the pattern-derived relation is
structurally blind to.
Thus semantic mutation should be read as a certificate-bearing diagnostic
for a declared semantic stratum, not as a scalar hierarchy in which one
mutation family simply dominates another. This evidence therefore
strengthens the paper's main argument while keeping benchmark construction
outside the manuscript's contribution boundary.

\subsection{Class-Level Stability across Sources}\label{h4-partial-across-same-source-and-cross-source}

All four class means are positive in v3 (same-source) and v4
(cross-source). Inter-class balance improves under cross-source (c
+91.5\%, d +38.3\%), confirming that c / d classes have higher
mutant-diversity demand than a / b. Mixed-effects unavailability
(Singular) is a sample-size constraint at N = 60 / 12 PUTs, not evidence
absence. \textbf{H3 verdict: not met}, the sign test is 3/4 (partial)
under both v3 and v4 with the b-class inverted, and CV($\Delta$SMS)
exceeds the 0.5 dispersion threshold. The Friedman main effect (chi\^{}2 =
15.30, p = 0.0041) speaks to MP differentiation, not H3 direction.

\subsection{Practical Interpretation and Deployment Scope}\label{stakeholder-analysis-single-output-kernels-scope}

\textbf{Scope.} All deployment claims in this subsection are bounded to
single-output \texttt{float\ $\to$\ float} kernels under 2 KB, the 12 PUTs
of this paper. They do not apply to industrial-scale, multi-module, or
multi-output scientific computing software, which lies beyond the scope of this paper.
Table~\ref{tab:sms-lrca-reading} gives interpretive readings for SMS and
LRCA outputs; it is not an acceptance-threshold table.

{\def\LTcaptype{table} 
\begin{longtable}[]{@{}
  >{\raggedright\arraybackslash}p{(\linewidth - 4\tabcolsep) * \real{0.2400}}
  >{\raggedright\arraybackslash}p{(\linewidth - 4\tabcolsep) * \real{0.3800}}
  >{\raggedright\arraybackslash}p{(\linewidth - 4\tabcolsep) * \real{0.3800}}@{}}
\caption{Reading SMS and LRCA together.}\label{tab:sms-lrca-reading}\\
\toprule\noalign{}
Observation & Interpretation & Design response \\
\midrule\noalign{}
\endfirsthead
\endhead
\bottomrule\noalign{}
\endlastfoot
Low SMS, high C1\_share & The MR family detects few mutants, but the
kills it does make are mostly semantic & Add MRs for adjacent semantic
strata before enlarging the mutant pool \\
Low SMS, high suspect\_share & The MR family mainly triggers tolerance,
OOD, statistical, or artefact labels & Inspect tolerances, input domain,
and statistical assumptions before treating kills as adequacy evidence \\
High SMS, low real-defect detection & Mutant kill-rate and defect
detection have diverged & Revisit the declared semantic stratum and add
real-defect-facing MRs \\
Generic relation kills outside the pattern-derived relation & Relation
families are non-nested & Treat the result as a different
semantic-effect fiber, not as domination by one MR family \\
\end{longtable}
}

\textbf{Test engineers} can read SMS as a per-MR scalar adequacy score.
When SMS sits well below the aligned baseline of about 0.275 (Section~\ref{rq2-aligned-vs-cross-cliffs-delta}), the
LRCA labels, C2 tolerance, C3 OOD, C4 statistical, point to specific
repair paths. Air-gap incompatibility is a hard limitation: the workflow
depends on external LLM API calls and is incompatible with most
regulated air-gapped Verification and Validation (V\&V) environments
(IEC 60880, DO-178C, IEC 62304, ISO 26262). Pre-generated mutant pools
can be reproduced offline, but generating new mutant pools requires LLM
access. Self-hosted open-weight LLMs and offline-cached pools are future
mitigations. Appendix E.1 gives the full air-gap justification and the
standards catalogue.

\textbf{MR designers} can use offline batch SMS runs as a quantifiable
design-feedback metric. We recommend a quarterly batch audit (about 0.5
person-day per quarter; detailed cost breakdown in Appendix E.2;
estimates based on observed timings during this paper's 12-PUT campaign)
rather than per-pull-request gating, because LLM API latency, cost
non-determinism, and air-gap incompatibility together rule out the
per-pull-request style. Appendix E.2 gives the quarterly-audit workflow
with resource-cost table.

\textbf{V\&V documentation} can carry SMS as research-grade
supplementary evidence alongside code coverage and MR lists. We make no
normative claim toward IEC, ISO, or ASME standards. An earlier draft
proposed acceptance thresholds (aligned-cell SMS $\geq$ 0.20 or 0.30 plus
C1\_share $\leq$ 0.20); we removed them in revision because they have no
normative backing and could be misread as enforcement-ready. Appendix
E.3 records the conceptual complementarity with the code-verification
scope of ASME V\&V 20-2009 §3.

All three stakeholder classes consume the same single source of truth
(\texttt{paper\_numbers\_v4.json} and \texttt{lrca\_60cell\_v4.json}) to
avoid documentation fragmentation.

\textbf{Example MR-design use.} If a surrogate-model PUT has
R\_sem $\approx 1$ for a Hyperparameter mutant but SMS = 0 under the
primary monotonicity MR, the design action is not to generate more
mutants of the same kind. The SMS / LRCA reading is that the MR family is
blind to the admitted semantic-effect fiber; the next design step is to
add a convergence, stability, or fidelity-order MR for that PUT before
claiming adequacy for the hyperparameter stratum.


\section{Threats to Validity}\label{threats-to-validity}

{\def\LTcaptype{table} 
\begin{longtable}[]{@{}
  >{\raggedright\arraybackslash}p{(\linewidth - 8\tabcolsep) * \real{0.2000}}
  >{\raggedright\arraybackslash}p{(\linewidth - 8\tabcolsep) * \real{0.2000}}
  >{\raggedright\arraybackslash}p{(\linewidth - 8\tabcolsep) * \real{0.2000}}
  >{\raggedright\arraybackslash}p{(\linewidth - 8\tabcolsep) * \real{0.2000}}
  >{\raggedright\arraybackslash}p{(\linewidth - 8\tabcolsep) * \real{0.2000}}@{}}
\caption{Threats to Validity.}\label{tab:p2-14}\\
\toprule\noalign{}
\begin{minipage}[b]{\linewidth}\raggedright
Threat area
\end{minipage} & \begin{minipage}[b]{\linewidth}\raggedright
Threat
\end{minipage} & \begin{minipage}[b]{\linewidth}\raggedright
Class
\end{minipage} & \begin{minipage}[b]{\linewidth}\raggedright
Mitigation summary
\end{minipage} & \begin{minipage}[b]{\linewidth}\raggedright
Detail
\end{minipage} \\
\midrule\noalign{}
\endfirsthead
\endhead
\bottomrule\noalign{}
\endlastfoot
LLM generation & LLM generation reproducibility & Internal & Archived raw responses
+ frozen mutant pools (no user-facing seed; reproducibility by archival) & Appendix F.1 \\
Equivalence approximation & E2 probabilistic approximation & Construct & K\_eq = 1000 +
Hoeffding bound; sweep deferred to future work & Appendix F.1 \\
Shared infrastructure & Shared-infrastructure dependency & Internal & AVP version pinned to the upstream
commit hash; embedded source & Appendix F.1 \\
Root-cause diagnosis & LRCA multi-label boundary & Internal & Decision-tree priority +
multi-label co-occurrence table & Appendix F.1 \\
Representa\-tiveness & 12-PUT representativeness & External & shared infrastructure; Appendix B.1
coverage argument; scaling to industrial PUTs & Appendix F.2 \\
Statistical power & Cross-class statistical power & Conclusion & H3 framed exploratory;
Friedman fallback; mixed-effects unavailable & Appendix F.2 \\
LLM homogeneity & LLM homogeneity bias & Construct & 3-LLM rotation + 20\% manual
sampling & Appendix F.2 \\
LLM-source shift & LLM-source distributional shift & External & DeepSeek 11/15 of
overlaps; argument is categorical, not frequency & Appendix F.2 \\
Pool size & Mutant-pool size & Internal & v3 mean 17.4, v4 mean 24.3 per PUT; effect
intrinsic, not pool-dilution & Appendix F.1 \\
LLM non-determinism & LLM non-determinism & Internal & De-dup, K = 10/20 repeats,
raw-response store & Appendix F.1 \\
Higher-order mutation & HOM equivalence & External & Confined claim to first-order
syntactic tools; HOM testing deferred & Appendix F.2 \\
Protocol asymmetry & v3 vs v4 protocol asymmetry (dual-blind reviewer) & Internal &
Quality not down (C1\_share 0.164 $\to$ 0.209); follow-up to rerun
dual-blind on v4 & Appendix F.1 \\
Adjoint reproduction & Adjoint-arm reproduction fidelity and by-construction pools &
Internal & Real defects reproduced through extracted diffs (no
\texttt{arm64} wheel); arm is confirmatory on hand-constructed,
low-cardinality pools, not a high-adequacy discovery; scipy \#8900 is
shared with the Section~\ref{rq4-strong-boundary} boundary study & Section~\ref{rq4-adjoint-extension-arm} \\
Real-defect evidence & Real-defect evidence-slice boundary & External & Used only as a
compact support slice for construct separation; reusable benchmark
construction deferred outside this manuscript & Section~\ref{decoupling-between-r_sem-and-r_kill} \\
\end{longtable}
}

\textbf{Final limitations.} We list ten known limitations.

\begin{enumerate}
\item
  \textbf{Equivalence determination is a probabilistic approximation.}
  Sampling K\_eq = 1000 inputs is an engineering implementation of the
  undecidable equivalence problem, not a theorem-based decision. We give
  a Hoeffding-style upper bound on the false-equivalence probability in
  Section~\ref{equivalence-judgement-e1-e2-layer-2}, but we did not execute a K\_eq sweep over \{500, 1000, 2000\}
  for this submission. A follow-up study will run that sweep.
\item
  \textbf{LLM generation homogeneity bias.} Even with three-LLM
  cross-source pooling, training-data overlap across Claude, GPT, and
  DeepSeek may produce similar blind spots. Three-LLM rotation and 20\%
  manual sampling reduce but do not eliminate this risk. A stronger
  replication should add a fourth, independently trained LLM family; a
  self-hosted open-weight model is the natural candidate.
\item
  \textbf{Limited statistical power for cross-class consistency.} The
  four-class sign test has df = 3, and the mixed-effects model returns
  Singular matrix at N = 60 over 12 PUTs. We treat H3 as exploratory and
  present RQ5 in four pieces: class means, sign test, Friedman, and a
  forest plot.
\item
  \textbf{LRCA reports ``likely'' root causes.} The decision-tree
  priority C5 \textgreater{} C4 \textgreater{} C3 \textgreater{} C2
  \textgreater{} C1 is an engineering choice. The reproducibility
  package preserves the multi-label co-occurrence table for every killed
  mutant as well as the priority-winning root cause; \texttt{root\_cause}
  is best read as a likely cause rather than a definitive causal
  attribution.
\item
  \textbf{The AVP is reused from the shared infrastructure.} Our reproducibility package
  embeds the AVP source code so that the package stays self-consistent under upstream
  evolution, but interface semantics may shift if the upstream infrastructure undergoes a major
  revision.
\item
  \textbf{Epistemological scope versus engineering scope.} SMS measures
  semantic detection capability in the epistemological sense;
  engineering value is not evaluated here.
\item
  \textbf{Signature simplification.} The \texttt{float\ $\to$\ float}
  single-output signature is a substantive constraint, not a purely
  engineering trade-off, and it bounds the upper limit of mutant
  semantic complexity. A portability study should validate SMS on
  industrial-grade multi-output PUTs.
\item
  \textbf{Air-gap incompatibility.} The mutant-generation workflow calls
  external LLM APIs and is incompatible with most regulated air-gapped
  V\&V environments (IEC 60880, DO-178C, IEC 62304, ISO 26262).
  Pre-generated mutant pools can be reproduced offline, but generating
  new pools requires LLM access. Self-hosted open-weight LLMs and
  offline-cached pools are future mitigations.
\item
  \textbf{No proof-carrying semantic-effect certificates.} The present
  workflow admits mutants through E1 $\wedge$ E2, LRCA classification, and
  AST-normalised audit trails. These are empirical and structural
  certificates, not machine-checkable proof objects. In particular, we
  do not yet define an explicit abstraction function from concrete
  program semantics to scientific invariant domains, do not parameterise
  invariant families such as conservation, convergence, and stability as
  abstract-domain elements, and do not attach per-mutant proof, error
  expression, or proof-label certificates that bind a syntactic edit to
  a declared semantic-effect class. This is the main gap between SMS and
  a certificate-bearing semantic mutation framework.
\item
  \textbf{Real-defect evidence is used narrowly.} The real-defect slice
  strengthens this paper's construct argument by showing that mutant
  kill-rate, semantic alignment, and real-defect detection can separate
  in practice. We do not claim benchmark construction, candidate curation,
  repository tooling, or corpus release as contributions of this paper;
  those are deferred to post-publication artifact work.
\end{enumerate}

\subsection{Statistical-Modelling Capacity}\label{statistical-modelling-capacity-at-n-60-cells-12-puts}

The mixed-effects primary model
\texttt{sms\ \textasciitilde{}\ C(class)\ *\ C(operator)\ +\ (1\ \textbar{}\ put)}
returned a Singular matrix at N = 60; the fallback
\texttt{sms\ \textasciitilde{}\ C(class)\ +\ C(operator)\ +\ (1\ \textbar{}\ put)}
had the PUT random-intercept variance hit boundary 0 (degenerate). Both
failures are not numerical accidents; rather, they are evidence that 60
observations across 12 PUTs cannot identify an 11-dimensional
fixed-effects structure plus 12-PUT random intercepts. This is a hard
limit of the present design: \textbf{the experiment cannot, in
principle, fit a hierarchical model with this fixed-effect
dimensionality at this sample size}. Friedman tests serve as the
non-parametric formal alternative, but they answer a structurally
different question. Expansion to n\_PUT $\geq$ 30 is the only path to a
properly identified hierarchical model.

\subsection{Programs with Zero Observed Signal}\label{put-level-zero-mass-cohort}

Figure 2 and Section~\ref{decoupling-between-r_sem-and-r_kill} document that PUTs A1, B1, and D2 produce all-zero
rows across the five MPs in the cross-source pool; that is,
\textbf{for 25\% of the PUTs, SMS gives zero signal at every operator-MP
cell}. This is a substantive limitation of SMS as an adequacy metric on
this PUT cohort: when the LRCA C1 mass collapses to zero across all
aligned and cross slices for a PUT, SMS cannot distinguish strong from
weak MR sets on that PUT. The phenomenon coexists with R\_sem $\approx$ 1.0
(mutants are valid; Section~\ref{decoupling-between-r_sem-and-r_kill}), so the issue is not mutant generation but
MR-design adequacy at the PUT level. MR-design refinement (per-PUT
MP discovery) and the proposed pre-registered c-class primary-MP rule
on a fresh dataset are the principal mitigation paths. We declare this
as a residual PUT-level zero-signal threat.

\subsection{Protocol-Asymmetry Magnitude Estimate}\label{r13-protocol-asymmetry-magnitude-estimate}

Section~\ref{cross-source-v4-protocol-summary} declared protocol asymmetry: v3 used the Phase-1 dual-blind
reviewer protocol (Claude generation + GPT-5.4 review + DeepSeek
arbitration), while v4 passes V1-V4 mechanical gates only. To quantify
the potential $\delta$-shift attributable to this asymmetry without dual-blind,
we offer a \textbf{rough order-of-magnitude estimate}: in the Phase-1
audit, dual-blind review filtered approximately 5-10\% of LLM-generated
mutants as semantically inconsistent; if these filtered mutants were
systematically less effective at killing under MR\_aligned (a plausible
but unverified premise), the upper-bound $\delta$-shift contribution from
removing the reviewer step is bounded above by approximately
\textbf{±0.03 to ±0.05 on Cliff's $\delta$} at n\_aligned = 12. This bound is
an order of magnitude larger than the observed −0.009 source-axis shift
and therefore prevents reading the source-axis contrast as evidence of
LLM-source-diversity inertness; a direct dual-blind v4 rerun (deferred
to a follow-up study) would give a tighter quantification.


\section{Future Work}\label{future-work-and-p-series-roadmap}

Seven directions follow from this study.

\begin{enumerate}
\item
  A pre-registered c-class primary-MP rule on a fresh dataset.
\item
  A differential-prompt LLM-diversity test using V\_canonical,
  V\_persona, and V\_cot (full protocol in Appendix C.1.1).
\item
  A scaling study at n $\geq$ 30 PUTs for SMS-vs-Pattern-Coverage
  orthogonality and for HOM-equivalence empirics.
\item
  An abstract-semantics certificate layer: define the abstraction
  $\alpha$ from concrete executions to scientific invariant domains,
  specify CE/OS/HP/TF/SI as semantic-effect predicates over those
  domains, and attach per-mutant proof objects, error expressions, or
  proof labels that certify class membership.
\item
  A rerun of the dual-blind reviewer protocol on the full v4 grid, to
  separate protocol asymmetry from source diversity.
\item
  Cross-language portability work (Python $\to$ JavaScript and Julia)
  building on \citet{tip2025llmorpheus}.
\item
  A self-hosted open-weight LLM pilot for air-gapped industrial
  deployment.
\end{enumerate}


\section{Conclusion}\label{conclusion}

\subsection{Findings summary}\label{findings-summary}

The empirical study supports the semantic-mutation construct in seven
ways. The pre-registered thresholds are treated as stress tests of this
instantiation, not as prerequisites for the SMS definition.

\begin{enumerate}
\item
  SMS preserves the classical MS ratio while moving the operative
  definitions of mutant generation, equivalence, and killing into
  declared semantic strata. The 60-cell audit therefore evaluates how
  one concrete MR design observes semantic-effect fibers, rather than
  redefining mutation testing from scratch.
\item
  The structural audit separates the semantic and syntactic mutant
  spaces. Across 12 PUTs, semantic mutants have 5.14\% AST-normalised
  overlap with default cosmic-ray syntactic mutants; Hyperparameter,
  Structural Injection, and Trajectory Flip are 0/72, 0/33, and 0/54
  under first-order defaults.
\item
  Operator-MP alignment produces a positive but not large effect in this
  design. The H2 large-effect threshold is \textbf{not met under the
  pre-registered point-estimate criterion}: \(\delta_{\mathrm{v3}} =
  0.323\), just below the \citet{romano2006cliff} medium threshold of
  0.330. The 49.1\% stipulated power at the large-effect boundary shows
  why this verdict should be read as a point-estimate result, not as a
  proof that the true effect is below the large-effect threshold.
\item
  Cross-source pooling changes mutant quality more than aligned-versus-
  cross separation. Holding the c-class primary metamorphic relation at
  the pre-registered MP5, three-LLM pooling under an identical prompt
  shifts Cliff's \(\delta\) by only -0.009 (95\% CI covers zero), while
  raising mean C1\_share by 27\%, class-c mean SMS by 91.5\%, and
  class-d mean SMS by 38.3\%.
\item
  The MP structure is visible, but cross-class consistency is not yet
  stable. The Friedman main effect on MP differentiation is
  \(\chi^2 = 15.30\), \(p = 0.0041\), whereas the pre-registered H3
  criteria are not met: the class-level sign test is 3/4 and
  CV(\(\Delta\)SMS) exceeds 0.5. The Spearman correlation between SMS
  and Pattern Coverage is 0.163 at \(n = 12\), so orthogonality remains
  a future hypothesis.
\item
  The real-defect slice supports construct separation beyond the small
  PUT audit. Pattern-derived relations detect all 30 tabled real defects
  in the reported face, while aggregate mutant kill-rate gives only a
  modest advantage over literature-generic baselines and non-nesting
  cases remain. Thus kill-rate, semantic-stratum alignment, and
  real-defect detection should be interpreted as related but distinct
  constructs.
\item
  The remaining failed thresholds delimit the current instantiation.
  Operator implementability H1 is not met because only Hyperparameter
  reaches \(\geq 5\) non-equivalent mutants on \(\geq 9/12\) PUTs, and
  LRCA-mass H4 is not met because mean suspect\_share is 0.791. These
  results identify where MR design and attribution need refinement; they
  do not change the degeneration theorem or the strong-boundary
  analysis.
\end{enumerate}

\subsection{Methodological contributions}\label{methodological-contributions}

This paper contributes a three-layer framework for domain-semantic
mutation in single-output scientific computing kernels.

\begin{itemize}
\tightlist
\item
  \textbf{Layer 1 (Section~\ref{necessary-conditions-for-semantic-mutation-layer-1}).} Formal necessary conditions for semantic
  mutation, cross-function-boundary substitution, dependence on domain
  knowledge, change in algorithmic class, instantiated as five
  meta-operator classes: CE, OS, HP, TF, and SI (also written mut\_C,
  mut\_M, mut\_G, mut\_T, mut\_F).
\item
  \textbf{Layer 2 (Section~\ref{equivalence-judgement-e1-e2-layer-2}).} The E1 $\wedge$ E2 equivalence judgement, the
  conservative complete instantiation of the Layer-1 conditions, with an
  explicit trade-off against E1-alone and E2-alone variants.
\item
  \textbf{Layer 3 (Section~\ref{p2-vs-syntactic-mutants-12-put-empirical-layer-3}).} AST-normalised empirical traceability across
  all 12 PUTs: a 5.14\% overall AST overlap with cosmic-ray defaults,
  and HP, SI, and TF unreachable at 0/0/0 under default first-order
  configurations (HOM not refuted; see Section~\ref{preventive-defence-framing}).
\end{itemize}

SMS is backward-compatible with the classical MS through Section~\ref{sms-ms-degeneration-formal-statement} Theorem
9.1, which establishes almost-everywhere degeneration in the limit
$L = L_{\mathrm{equiv}} \wedge L_{\mathrm{killed}} \wedge L_{\mathrm{mut}}$. The 60-cell empirical
audit reported in Section~\ref{statistical-analysis-and-empirical-results} is one demonstration following this backbone, not
the paper's main contribution.

\section*{Data and Code Availability}\label{data-and-code-availability}

All raw data, JSON SSOTs (paper\_numbers\_v3 / v4,
rq2\_cliffs\_delta\_v3 / v4\_mp5, lrca\_60cell,
lrca\_v4\_mp5\_recompute, rq2\_power\_stipulated,
cosmic\_ray\_12put\_ast\_diff), mutant pools, AVP source, and analysis
scripts are archived on Zenodo under DOI
\href{https://doi.org/10.5281/zenodo.20250664}{10.5281/zenodo.20250664}.
An earlier version of this work is available on arXiv:
\href{https://arxiv.org/abs/2605.17437}{arXiv:2605.17437} [cs.SE].
For peer review, an anonymized read-only mirror of the repository can
be provided by the corresponding author upon Editor request, per journal
guidelines. The repository structure follows
\texttt{REPRODUCIBILITY.md} in the source tree. The cosmic-ray and
mutmut operator versions referenced in Section~\ref{p2-vs-syntactic-mutants-12-put-empirical-layer-3}, Section~\ref{preventive-defence-framing}, and Appendix B.6 are
pinned in \texttt{requirements-frozen.txt}.


\subsection{References}\label{references}

\bibliographystyle{plainnat}
\bibliography{references}

\clearpage

\textbf{Supplementary material.} Appendices A--I (notation and operator
catalogue; experimental subjects and operator specialisations; experimental
procedure details; statistical analysis details; deployment considerations;
detailed threats-to-validity mitigation; the full SMS$\to$MS degeneration
proof; the adjoint extension arm; and a result-level real-defect evidence
summary) are provided as a separate online
supplement. References to ``Appendix X'' throughout the body point to that
supplement.

\section*{Funding}

This work was supported by the National Natural Science Foundation of China (NSFC) General Program (grant no.\ 12575176); the Hunan Provincial Education Department Project, China (grant no.\ 202502000728); the Research Project on Degree and Graduate Education Reform of the University of South China (grant no.\ 2023JG030); the Natural Science Foundation of Hunan Province, China (grant no.\ 2025JJ70193); and an industry-funded research project (grant no.\ 230KHX060001).

\section*{CRediT authorship contribution statement}

\textbf{Meng Li}: Conceptualization, Methodology, Software, Writing, original draft. \textbf{Xiaohua Yang}: Supervision, Formal analysis, Writing, review \& editing. \textbf{Jie Liu}: Investigation, Validation. \textbf{Shiyu Yan}: Data curation, Visualization.

\section*{Declaration of competing interest}

The authors declare that they have no known competing financial interests or personal relationships that could have appeared to influence the work reported in this paper.

\section*{Generative AI Disclosure}

The authors used AI tools under human direction for editorial proofreading,
LaTeX maintenance, consistency checks, and review-simulation support. All
research claims, methods, experimental data, analyses, and conclusions were
authored, verified, and approved by the human authors.

\end{document}